# Computer-Aided Extraction of Select MRI Markers of Cerebral Small Vessel Disease: A Systematic Review

Jiyang Jiang [1,*], Dadong Wang [2], Yang Song [3], Perminder S. Sachdev [1,4], Wei Wen [1,4]

1. Centre for Healthy Brain Ageing, Discipline of Psychiatry and Mental Health, School of Clinical Medicine, Faculty of Medicine, University of New South Wales, NSW 2052, Australia.
2. Quantitative Imaging Research Team, Data61, CSIRO, Marsfield, NSW 2122, Australia.
3. School of Computer Science and Engineering, University of New South Wales, NSW 2052, Australia.
4. Neuropsychiatric Institute, Prince of Wales Hospital, Randwick, NSW 2031, Australia.

* Corresponding author: Level 1, Centre for Healthy Brain Ageing, AGSM building, UNSW Sydney, NSW 2052, Australia. E-mail address: jiyang.jiang@unsw.edu.au (J. Jiang).

## Abstract

Cerebral small vessel disease (CSVD) is a major vascular contributor to cognitive impairment in ageing, including dementias. Imaging remains the most promising method for in vivo studies of CSVD. To replace the subjective and laborious visual rating approaches, emerging studies have applied state-of-the-art artificial intelligence to extract imaging biomarkers of CSVD from MRI scans. We aimed to summarise published computer-aided methods to examine three imaging biomarkers of CSVD, namely cerebral microbleeds (CMB), dilated perivascular spaces (PVS), and lacunes of presumed vascular origin. Seventy-one classical image processing, classical machine learning, and deep learning studies were identified. CMB and PVS have been better studied, compared to lacunes. While good performance metrics have been achieved in local test datasets, there have not been generalisable pipelines validated in different research or clinical cohorts. Transfer learning and weak supervision techniques have been applied to accommodate the limitations in training data. Future studies could consider pooling data from multiple sources to increase diversity, and validating the performance of the methods using both image processing metrics and associations with clinical measures.

# 1 Introduction

Cerebral small vessel disease (CSVD) is highly prevalent in older adults, and is associated with cognitive and neurobehavioural symptoms (Verdelho *et al*., 2021). Imaging biomarkers of CSVD can include recent small subcortical infarct (RSSI), white matter hyperintensity (WMH) of presumed vascular origin, cerebral microbleed (CMB), dilated perivascular space (PVS), and lacune of presumed vascular origin (hereafter, lacune) (Wardlaw *et al*., 2013). Brain atrophy is also frequently seen in CSVD (Wardlaw *et al*., 2013), but the location and extent are usually comparable with those in ageing and neurodegenerative diseases, making it an unspecific biomarker. WMHs are CSVD lesions in brain white matter that appear abnormally bright on T2-weighted MRI scans. As a lesion with variable sizes and usually widely spread in the brain, WMH has been well documented for its automated segmentation methods (Balakrishnan *et al*., 2021; Caligiuri *et al*., 2015), risk factors (Brown *et al*., 2021), and clinical importance (Debette and Markus, 2010). RSSIs are infarcts in the perfusion territory of arterioles with imaging features or clinical symptoms consistent with a recent lesion occurring in the previous few weeks (Wardlaw *et al*., 2013). They usually evolve into WMHs or lacunes, or, more rarely, disappear and leave normal-appearing brain tissues on MRI. RSSIs appear hyperintense on T2-weighted fluid-attenuated inversion recovery (FLAIR) sequences, and hypointense on T1-weighted scans, similar to the contrast of WMH. Most WMH segmentation methods do not explicitly exclude RSSIs. Diffusion-weighted imaging can aid the identification of RSSIs. While studies have proposed methods to detect, segment and classify stroke lesions (Sarmento *et al*., 2020), as well as to separate them from WMH (Guerrero *et al*., 2018), there has not been any study aiming at explicitly segmenting RSSIs.

Unlike WMH, CMB, PVS and lacunes are focal lesions of small sizes (up to 15 mm in diameter). They share a few characteristics on MRI scans (Table 1): 1) focal lesions with darker or brighter intensities compared to surrounding tissues, and 2) round/ovoid (perpendicular plane for PVS) shape. While labour-intensive, error-prone, and subjective visual ratings by experts have been applied to the majority of previous studies, more recent studies have proposed various algorithms to automate the extraction of CMB, PVS and lacunes. However, these methodological studies aimed at proof of concept with limited local data for validation. There have not been any generally accepted, automated pipelines ready for clinical studies of CMB, PVS or lacunes. Such pipelines are essential to advance our understanding on CSVD, especially in the era of big public datasets (Madan, 2021). The current work systematically reviewed these methodological studies, critically appraised the algorithms from the perspective of generalisability, and highlighted promising techniques for future development, aiming at discovering potential future steps to bridge the gap towards practical and generalisable applications in both research and clinical settings. Since there has been a previous systematic review of PVS segmentation (Hernandez Mdel *et al*., 2013), we focused on studies published after 2013 for PVS.

A few terms used in this review need to be clarified. The current review included all automated or semi-automated studies trying to address *regression*, *detection* and/or *segmentation* tasks. Regression tasks were defined as the prediction of an overall scan-level measure, such as the rating or number of lesions. Detection tried to find lesions without explicitly outlining the boundary between the lesion and surrounding normal tissues. The output was usually a bounding box enclosing the lesion in detection tasks. Segmentation was defined as a voxel-by-voxel depiction of the lesion. Moreover, the applied techniques were categorised into *classical image* *processing*, *classical machine learning*, and *deep learning*. Classical image processing techniques referred to traditional computer vision methods without a learning component. In the context of this work, these included various transformation-based methods and intensity-based segmentation techniques. Classical machine learning techniques mainly included random forest (RF) and support vector machine (SVM) classifiers in this review. Deep learning techniques referred to convolutional neural networks (CNN) and their variants.

# 2 Methods

## 2.1 Identification of studies

Because of the technical nature of the review topic, two separate literature searches were conducted in PubMed and IEEE Xplore databases (November, 2021). Specifically, the following keywords were used: *microbleed*, perivascular space*, Virchow-Robin space*, lacune*, lacunar infarct*, lacunar stroke, brain infarct, detect*, segment*, extract*, identif*, quantif* (Figure 1). Reference lists of all relevant articles were also searched for additional studies.

## 2.2 Inclusion/exclusion criteria

For the systematic review, inclusion criteria were as follows:

- Methodological studies on the classification, detection, and/or segmentation of CMB, PVS and lacunes, defined in the Standards for Reporting Vascular Changes on Neuroimaging (STRIVE) consensus (Wardlaw et al., 2013).
- Studies using human MRI data for training, validation, and testing.

- Studies were required to apply computer-aided methods in a significant proportion of the algorithm.

Exclusion criteria were as follows:

- Development and/or comparison of imaging sequences.
- Studies using visual detection/rating methods.
- Studies not published in English.

# 3 Results

## 3.1 Cerebral microbleeds

### 3.1.1 Overview

Table 2 summarised studies of CMB. Many approaches could be divided into two stages: 1) *candidate detection* to screen for CMB candidates with high sensitivity, and 2) *false positive (FP) reduction* to remove false positives from the candidates. In the candidate detection stage, studies have used methods ranging from intensity thresholding (Barnes *et al*., 2011; Dou *et al*., 2015; Ghafaryasl *et al*., 2012), identifying circular/spherical objects (Bian *et al*., 2013; Chesebro *et al*., 2021; Fazlollahi *et al*., 2015; Fazlollahi *et al*., 2014; Morrison *et al*., 2018), to convolutional neural networks (Al-Masni *et al*., 2020; Dou *et al*., 2016). The FP reduction stage usually employs classical machine learning (Barnes *et al*., 2011; Chen *et al*., 2015; Dou *et al*., 2015; Dou *et al*., 2016; Fazlollahi *et al*., 2015; Fazlollahi *et al*., 2014; Ghafaryasl *et al*., 2012; Roy *et al*., 2015; van den Heuvel *et al*., 2015; van den Heuvel *et al*., 2016) and deep learning (Al-Masni *et al*., 2020; Liu *et al*., 2019) algorithms. Some deep learning (e.g., (Rashid *et al*., 2021)) and classical image processing (e.g., (Kuijf *et al*., 2013)) studies addressed the CMB segmentation problem in a single-pass fashion.

### 3.1.2 Materials

*Study samples* – The included studies had a median sample size of 27 participants. Due to the low prevalence of CMB in general population, studies tended to use diseased patients who were more prone to CMB. These included patients with Cerebral Autosomal Dominant Arteriopathy with Sub-cortical Infarcts and Leukoencephalopathy (CADASIL) (Hong *et al*., 2019; Hong *et al*., 2020; Wang *et al*., 2017; Wang *et al*., 2019; Zhang *et al*., 2018a; Zhang *et al*., 2018b; Zhang *et al*., 2016b), stroke/transient ischemic attack (Dou *et al*., 2015; Dou *et al*., 2016; Seghier *et al*., 2011), traumatic brain injury (Liu *et al*., 2019; Roy *et al*., 2015; van den Heuvel *et al*., 2015), and mild cognitive impairment/Alzheimer's Disease (Barnes *et al*., 2011; Fazlollahi *et al*., 2015; Fazlollahi *et al*., 2014), as well as individuals diagnosed with gliomas and had undergone radiation therapy that induced CMB (Bian *et al*., 2013; Chen *et al*., 2019; Morrison *et al*., 2018).

*MRI modalities* – Hemosiderin deposits in CMB are paramagnetic, leading to heterogenic distribution of magnetic field and quick decay of magnetic resonance signal around CMB (i.e., susceptibility effects) (Shams *et al*., 2015). T2*-weighted gradient recalled echo (GRE) and susceptibility-weighted imaging (SWI) are the most commonly used imaging modalities for the segmentation of CMB. The T2*-weighted GRE sequence is sensitive to susceptibility effects, and has been used to detect CMB for a long time (Greenberg *et al*., 2009).

SWI sequence is introduced later as an alternative to T2*-weighted GRE, because it is heavily weighted by the intrinsic tissue magnetic susceptibility (Ruetten *et al*., 2019; Shams *et al*., 2015), and results in greater reliability and sensitivity in detecting CMB (Cheng *et al*., 2013; Shams *et al*., 2015). However, this is at the cost of increased conspicuity of CMB mimics, including calcium and iron deposits, arteries, and veins (Cheng *et al*., 2013; Shams *et al*., 2015). Moreover, due to the greater blooming effects of SWI on paramagnetic substances, CMB tend to be more irregular shaped on SWI (Chesebro *et al*., 2021; Mittal *et al*., 2009).

Phase images can help separate diamagnetic substances (e.g., calcification) from paramagnetic CMB (Liu *et al*., 2019). Similarly, quantitative susceptibility mapping (QSM), a further advancement of the phase information to quantify tissue susceptibility (Ruetten *et al*., 2019), has also been used for the same purpose (Rashid *et al*., 2021).

*MRI preprocessing* – MRI preprocessing steps before CMB segmentation were straightforward, mainly including bias field correction (Ghafaryasl *et al*., 2012; Rashid *et al*., 2021; van den Heuvel *et al*., 2016), brain extraction (Al-Masni *et al*., 2020; Ghafaryasl *et al*., 2012; Kuijf *et al*., 2013; Morrison *et al*., 2018; van den Heuvel *et al*., 2016), and intensity normalization (Al-Masni *et al*., 2020; Chen *et al*., 2015; Fazlollahi *et al*., 2015; Kuijf *et al*., 2013). Some studies also implemented data interpolation (Barnes *et al*., 2011; Chesebro *et al*., 2021).

*Defining true CMB* – The definition of gold-standard CMB in most studies is based on the Microbleed Anatomical Rating Scale (MARS) (Gregoire *et al*., 2009), where definite CMB are defined as small, rounded or circular, well-defined hypointense lesions within brain tissue with a diameter of 2-10 mm. Some computer-aided studies (Bian *et al*., 2013) also classified possible CMB which were less rounded or circular, less dark, or difficult to distinguish from mimics with confidence (Bian *et al*., 2013; Gregoire *et al*., 2009). With dual echo SWI, a missing blooming effect would indicate a non-CMB (Kuijf *et al*., 2011). Brain Observer MicroBleed Scale (BOMBS) was also used in some studies for defining true CMB (Rashid *et al*., 2021).

### 3.1.3 Algorithms

#### 3.1.3.1 Classical image processing

Studies using classical image processing methods to extract CMB usually considered intensity and geometric information.

##### 3.1.3.1.1 Intensity

Watershed is an image processing algorithm mainly used for segmentation. The intuitive idea comes from geography where catchment basins are filled up with water (Roerdink and Meijster, 2000). When the water level has reached the highest point where water coming from different basins would meet, dams are built. These dams are called watershed lines or watershed. After Contrast Limited Adaptive Histogram Equalization (CLAHE), Bottom-Hat filtering, and K-means clustering, watershed was applied with active contour segmentation to segment CMB (Tajudin *et al*., 2017a; Tajudin *et al*., 2017b).

Seghier *et al*. (Seghier *et al*., 2011) considered CMB as an extra tissue type in the unified normalisation-segmentation implemented in SPM (Statistical Parametric Mapping). Briefly, priors were adjusted to accommodate the intensity distribution in T2*-weighted images. The contrast between different tissue types was also standardised. These empirical priors were applied in the initial unified normalisation-segmentation. The results were used to modify the priors, including CMB probability thresholding, morphological granulometry thresholding, and removing skull and cerebrospinal fluid regions. The refined priors were then used in the second iteration of unified normalisation-segmentation.

Liu *et al*. (Liu *et al*., 2020b) considered both intensity at the voxel level, and gradients obtained by applying first-order horizontal and vertical differential operators to the images. The Fourier descriptor was employed to capture shape information.

###### 3.1.3.1.2 Geometry

The 'blooming' effect of CMB in all directions on SWI, influenced by the echo time, makes it a good fit for the radial symmetry transform (RST; (Bian *et al*., 2013; Kuijf *et al*., 2012)) (Figure 2). In RST, Sobel filters, either 2D (Bian *et al*., 2013; Morrison *et al*., 2018) or 3D (Kuijf *et al*., 2011; Kuijf *et al*., 2013; Kuijf *et al*., 2012), were applied to SWI data, resulting in positive values for dark-to-bright gradients. Since CMB are hypointense on SWI, only negatively affected voxels should be considered for gradients pointing towards the center of CMB. Orientation and magnitude projection images were then calculated (Bian *et al*., 2013; Kuijf *et al*., 2012), followed by the application of RST for each radius $n \in N$. The final transform was computed as the sum of all radial symmetry contributions. The output from RST was a voxel-by-voxel map with intensities indicating local sphericalness (Kuijf *et al*., 2013). The center of CMB would have high intensities (i.e., high radial symmetry values) on the map.

Chesebro *et al*. (Chesebro *et al*., 2021) applied Canny edge detection on the 2D gradient image (derived from applying Sobel filter to SWI or T2*-weighted GRE), followed by a circular Hough transform to detect circles (Figure 2). After excluding circles located on the edge of the brain, within cerebrospinal fluid, or with unrealistic sizes, the remaining circles were output as CMB candidates. Four metrics for each candidate CMB region were then used to clean FPs: 3D image entropy, 2D image entropy of the maximum intensity projection, and the volume and compactness of the central blob resulting from Frangi filter. Moreover, Liu *et al*. (Liu *et al*., 2020b) used Fourier descriptors to capture shape information.

##### 3.1.3.2 Classical machine learning

SVM (Barnes *et al*., 2011; Dou *et al*., 2015) and RF (Dou *et al*., 2015; Fazlollahi *et al*., 2015; Fazlollahi *et al*., 2014; Roy *et al*., 2015; van den Heuvel *et al*., 2016) were the most commonly used machine learning algorithms in detecting CMB. Intensity, size and shape features were usually considered for the classification of CMB vs. non-CMB.

###### 3.1.3.2.1 Intensity and size features

Intensity features considered in classical machine learning studies of CMB usually included average, minimum, maximum, median and/or standard deviation of intensities in the candidate (Barnes *et al*., 2011; van den Heuvel *et al*., 2016). Some studies also considered intensities of the voxels surrounding the candidate to take into account the contrast of CMB with surrounding voxels (Ghafaryasl *et al*., 2012). CMB usually have a diameter of 2-10 mm (Gregoire *et al*., 2009; Wardlaw *et al*., 2013). Therefore, size features, including size of the candidate and the bounding box, were also considered (Ghafaryasl *et al*., 2012).

#### 3.1.3.2.2 Geometric features

Shape is an important descriptor for detecting CMBs and separating true CMBs from mimic. Unlike the straightforward intensity and size features, geometric features usually require sophisticated transforms.

Fazlollahi *et al*. (Fazlollahi *et al*., 2015; Fazlollahi *et al*., 2014) applied Radon transform to describe the shape of candidate objects and distinguish CMBs from vessel mimics (Figure 2). Specifically, the mean and standard deviation profiles across different projection angles were used as shape features. The length of mean and standard deviation profiles was standardized to account for different sizes of candidates.

The Hessian matrix and resultant measures have been used in both candidate detection and FP reduction stages. In the candidate detection stage, after applying multi-scale Laplacian of Gaussian filters, Fazlollahi *et al*. screened the resultant data for spherical objects (Fazlollahi *et al*., 2015; Fazlollahi *et al*., 2014). This was done through 1) transforming the question of detecting a 3D object into a question of detecting lines in orthogonal dimensions, 2) identifying the center of the object by averaging the normalized line responses from the previous step and extracting local maxima, and 3) assessing the anisotropy of eigenvalues from the Hessian matrix of the identified center for the detection of spherical objects. A lower anisotropy threshold was applied to account for the partial volume effects and presence of noise. The results were used as candidates for further reduction of FPs.

In the FP reduction stage, the three eigenvalues of the Hessian matrix and different synthetic measures from these eigenvalues have been proposed as shape descriptors:

- Ratio of eigenvalues (Ghafaryasl *et al*., 2012)
- Sphericalness defined as the volume to largest cross-sectional area (Fazlollahi *et al*., 2015; Fazlollahi *et al*., 2014).
- Largest cross section defined as the ratio of the largest 2 eigenvalues (Fazlollahi *et al*., 2015; Fazlollahi *et al*., 2014)
- Fractional anisotropy (Fazlollahi *et al*., 2015; Fazlollahi *et al*., 2014)
- Determinant of eigenvalues (Ghafaryasl *et al*., 2012; van den Heuvel *et al*., 2016)
- Trace defined as the sum of the diagonal in the Hessian matrix (Ghafaryasl *et al*., 2012; van den Heuvel *et al*., 2016)
- Frangi vesselness (Ghafaryasl *et al*., 2012; van den Heuvel *et al*., 2016)
- Local sphericalness defined as the equalness of 2 principal curvatures (van den Heuvel *et al*., 2015; van den Heuvel *et al*., 2016).
- To distinguish the distorted (e.g., elongated) CMB in SWI from vessels, a feature to describe the candidates' orientation, defined by the angle between the eigenvector with the largest eigenvalue and the xy-plane, was also employed (Fazlollahi *et al*., 2015; Fazlollahi *et al*., 2014).

A few other geometric descriptors have been proposed to describe CMB. van den Heuvel *et al*. (van den Heuvel *et al*., 2015; van den Heuvel *et al*., 2016) used the convolution outcome of the scan with a spherical kernel as a shape feature. The kernel was designed to be negative in the middle, and positive on the edge, resulting in positive response when convolving with a dark spherical object. Since the kernel had been normalized, there is no response when convolving with an image with constant intensities.

In addition, as introduced in the Section 3.1.3.1 'Classical image processing methods', RST outputs voxel-wise maps with intensities indicating local radial symmetry, and can be therefore used as shape features in classical machine learning approaches (Roy *et al*., 2015). The three eigenvalues of the covariance matrix and relative anisotropy calculated from these eigenvalues were considered as shape features in a previous study (Barnes *et al*., 2011). Moreover, the compactness, defined as the ratio between the volume of candidates and the volume of their bounding boxes, was used (Ghafaryasl *et al*., 2012).

#### 3.1.3.3 Deep learning

CNNs have been widely used in recent studies of CMB extraction (Table 2). Rashid *et al*. (Rashid *et al*., 2021) applied U-Net with a few modifications to adapt to characteristics in segmenting CMB. These modifications included 1) padded, instead of unpadded, convolutions, and 2) a higher number of resolution layers (6 as opposed to 5). 3D CNN was used in a few studies to take into account the contextual information in adjacent slices (Chen *et al*., 2019; Dou *et al*., 2016). Some recent studies also applied transfer learning techniques to address the issue of limited training data (Afzal *et al*., 2022; Hong *et al*., 2019; Li *et al*., 2021). Zhang *et al*. (Zhang *et al*., 2018b; Zhang *et al*., 2016b) introduced sparse autoencoder layers to the CNN model for CMB segmentation.

#### 3.1.3.4 CMB mimics

Vessel mimics are one of the most common false positives in detecting CMB. In addition to the shape features trying to distinguish vessels from CMB (e.g., Radon transform, vesselness using Hessian matrix), features from other imaging modalities have also been used. For example, Ghafaryasl *et al*. (Ghafaryasl *et al*., 2012) extracted average, minimum, maximum, median, and standard deviation of intensities within candidates on proton density (PD)-weighted images. Blood vessels are shown as hypointensity on PD-weighted scans, while CMB are not visible on these images. The phase information (either high-pass filtered phase images (Liu *et al*., 2019) or the calculated QSM images (Rashid *et al*., 2021)) was used to exclude diamagnetic calcification mimics.

#### 3.1.3.5 Limited training data and class imbalance

Under-sampling (e.g., (Zhang *et al*., 2016b)) and augmentation (e.g., (Afzal *et al*., 2022)) were the most common approach for the problems of class imbalance and small sample sizes. Dou *et al*. (Dou *et al*., 2015) applied a stacked 2-layer convolutional independent subspace analysis (ISA) network to find discriminative features of CMB from limited training data in an unsupervised fashion, which were then used in a SVM classifier. Transfer learning with pre-trained models was also used in some recent deep learning studies to address the issue of limited training data. ResNet50 (Afzal et al., 2022; Hong et al., 2019), AlexNet (Afzal et al., 2022), and VGG-based Single Shot Multi-box Detector (SSD)-512 network (Li et al., 2021) pre-trained on the ImageNet dataset have been used to initialise the optimisation.

After the initial candidate detection, Fazlollahi *et al*. (Fazlollahi *et al*., 2015) used consecutive independent RF classifiers with bagging and random undersampling. Low to high posterior probability thresholds were applied to the output from the classifiers to progressively exclude obvious non-CMB candidates, to handle the imbalanced dataset (i.e., more non-CMBs than non-CMBs) and decrease the number of non-informative candidates without distinctive features. These techniques have proven to be beneficial for effectively reducing FPs.

## 3.2 Perivascular spaces

### 3.2.1 Overview

Table 3 summarised studies of PVS. Out of the 25 studies, 9 studies employed classical image processing methods (Ballerini *et al*., 2020; Ballerini *et al*., 2016; Ballerini *et al*., 2018; Boespflug *et al*., 2018; Descombes *et al*., 2004; Liu *et al*., 2020a; Schwartz *et al*., 2019; Sepehrband *et al*., 2019; Wang *et al*., 2016), 5 applied classical machine learning techniques (SVM (González-Castro *et al*., 2016; González-Castro *et al*., 2017) and RF (Park *et al*., 2016; Zhang *et al*., 2016a, 2017)), 1 study combined classical image processing and machine learning (RF, (Hou *et al*., 2017)), and 10 used deep learning models (Boutinaud *et al*., 2021; Dubost *et al*., 2019a; Dubost *et al*., 2020; Dubost *et al*., 2017; Dubost *et al*., 2019b; Jung *et al*., 2019; Lian *et al*., 2018a; Lian *et al*., 2018b; Sudre *et al*., 2019; Yang *et al*., 2021). Due to the high prevalence and number of PVS within a scan, studies usually focus on certain brain regions, especially basal ganglia (Boutinaud *et al*., 2021; Dubost *et al*., 2019a; Dubost *et al*., 2020; Dubost *et al*., 2017; Dubost *et al*., 2019b; González-Castro *et al*., 2016; González-Castro *et al*., 2017; Wang *et al*., 2016; Yang *et al*., 2021) and centrum semiovale (Ballerini *et al*., 2020; Ballerini *et al*., 2016; Ballerini *et al*., 2018; Dubost *et al*., 2020; Dubost *et al*., 2019b; Schwartz *et al*., 2019), because arterioles in these regions are relatively straight appearing as straight lines or perforating through the imaging plane (Wang *et al*., 2016). Midbrain, hippocampi, and deep corona radiata were also included in some studies (Ballerini *et al*., 2020; Dubost *et al*., 2019a; Dubost *et al*., 2020).

There were also studies including wider regions, such as white matter (Boutinaud *et al*., 2021; Liu *et al*., 2020a) and the whole brain (Boespflug *et al*., 2018; Descombes *et al*., 2004; Lian *et al*., 2018a; Lian *et al*., 2018b; Park *et al*., 2016; Sepehrband *et al*., 2019; Sudre *et al*., 2019). Studies have treated the task as a regression (rating scores (Yang *et al*., 2021), number of PVS (Dubost *et al*., 2019a; Dubost *et al*., 2019b)), detection (Dubost *et al*., 2020; Dubost *et al*., 2017; Sudre *et al*., 2019), or segmentation (Ballerini *et al*., 2020; Ballerini *et al*., 2016; Ballerini *et al*., 2018; Boespflug *et al*., 2018; Boutinaud *et al*., 2021; Descombes *et al*., 2004; González-Castro *et al*., 2016; González-Castro *et al*., 2017; Lian *et al*., 2018a; Lian *et al*., 2018b; Liu *et al*., 2020a; Park *et al*., 2016; Schwartz *et al*., 2019; Wang *et al*., 2016; Zhang *et al*., 2016a, 2017) question.

A few studies aimed to enhance MRI data for PVS segmentation (Hou *et al*., 2017; Jung *et al*., 2019; Sepehrband *et al*., 2019).

### 3.2.2 Materials

*Study samples* – Studies have been using population-based cohorts (Ballerini *et al*., 2020; Ballerini *et al*., 2016; Boespflug *et al*., 2018; Dubost *et al*., 2019a; Dubost *et al*., 2017; Dubost *et al*., 2019b; Schwartz *et al*., 2019), young healthy individuals (Boutinaud *et al*., 2021; Hou *et al*., 2017; Jung *et al*., 2019; Lian *et al*., 2018a; Lian *et al*., 2018b; Park *et al*., 2016; Sepehrband *et al*., 2019), and patients with subjective memory decline/mild cognitive decline/dementia (Ballerini *et al*., 2018; Schwartz *et al*., 2019; Yang *et al*., 2021), stroke (González-Castro *et al*., 2016; González-Castro *et al*., 2017; Wang *et al*., 2016), and idiopathic generalised epilepsy (Liu *et al*., 2020a). The median sample size is 65 (range, 1-2202).

*MRI modalities* – All studies used T1-weighted, T2-weighted, and/or T2-weighted FLAIR sequences for extracting PVS, except for 3 studies that employed proton density-weighted MRI with similar contrast to T2-weighted images (Boespflug *et al*., 2018; Dubost *et al*., 2019a; Dubost *et al*., 2017). A few studies also tried to enhance contrast (Hou *et al*., 2017; Jung *et al*., 2019) and segment (Lian *et al*., 2018a; Lian *et al*., 2018b; Park *et al*., 2016; Zhang *et al*., 2016a, 2017) PVS from scans acquired from 7T MRI scanners.

*MRI preprocessing* – Since a few computer-aided methods for extracting PVS focused on particular brain regions (e.g., basal ganglia, centrum semiovale), these masks were generated in addition to the preprocessing steps commonly used in studies of CMB and lacunes (within-subject coregistration of image modalities, bias field correction, brain extraction, intensity normalisation). Subcortical grey matter structures segmented in FreeSurfer had been used to derive masks of the basal ganglia which included thalamus, thalamus-proper, caudate, putamen, pallidum, and accumbens (Boutinaud *et al*., 2021). Other methods had detected landmarks in the insular cortex to construct a polygon around the basal ganglia region (González-Castro *et al*., 2016; González-Castro *et al*., 2017). Cerebrospinal fluid regions were then removed from the polygon to generate a mask of the basal ganglia. Centrum semiovale was usually defined as white matter regions above the lateral ventricles (Ballerini *et al*., 2016). Some studies registered a centrum semiovale mask in standard space to individual space (Ballerini *et al*., 2020). Others used FreeSurfer segmentations (Dubost *et al*., 2019b), or first detected lateral ventricles and then defined white matter above them as centrum semiovale (Ballerini *et al*., 2018).

*Defining true PVS* – While studies applied various protocols to the manual annotations and visual ratings of PVS (Ballerini *et al*., 2020; Boespflug *et al*., 2018; Boutinaud *et al*., 2021; Dubost *et al*., 2020), PVS are usually defined as round, ovoid, or linear lesions (diameter < 3mm) with intensity characteristics similar to cerebrospinal fluid. The pattern of PVS in different brain regions may show slight differences with hippocampal PVS being rounder, and PVS in centrum semiovale being more elongated (Dubost *et al*., 2019b). As PVS follow the trajectory of a vessel, the PVS shape on a scan slice depends on the plane (Wardlaw *et al*., 2013). Some classical machine learning studies applied semi-automated methods followed by manual correction to accelerate the annotation of true PVS (Park *et al*., 2016; Zhang *et al*., 2016a, 2017). Generating gold standard for the image enhancement task in PVS segmentation is challenging. A deep learning study (Jung *et al*., 2019) used established classical image processing methods (Hou *et al*., 2017) to generate gold standard data and validate its proposed model.

### 3.2.3 Algorithms

#### 3.2.3.1 Classical image processing

##### 3.2.3.1.1 Intensity

Wang *et al*. (Wang *et al*., 2016) applied linear intensity adjustment to the basal ganglia slice on the T2-weighted scan. The voxel-by-voxel product of the original T2 scan with this intensity adjusted T2 image was then calculated. The basal ganglia region was then segmented. After interactive adjustment of minimum and maximum sizes of hyperintense objects, one of three intensity thresholds was applied. The selection of threshold was based on the severity of PVS, background intensity, and other visible lesions.

In another study, voxel-level intensity inhomogeneity was used to segment PVS candidates (Schwartz *et al*., 2019). Specifically, the median intensity of surrounding voxels within a search filed of radius of 3.5 mm was defined as the median score for the voxel i. The mean intensity differences between the voxel i and its neighbouring voxels within a search field of radius of 5.5 mm was defined as the difference score for voxel i. PVS voxels were then defined as those within the white matter mask, with an intensity no more than 90% of the median score, and a difference score larger than 5% of their intensities. The thresholding was done on T1-weighted images, but taking into account T2-weighted FLAIR data to avoid WMH regions. Geometric and size constraints were then applied to reduce FPs.

#### 3.2.3.1.2 Geometry

Provided that the pattern of PVS follows the course of blood vessels, filters sensitive to vessel-like tubular structures will be helpful in enhancing the presence of PVS in images (Figure 2). Liu *et al*. (Liu *et al*., 2020a) and a body of work by Ballerini *et al*. (Ballerini *et al*., 2020; Ballerini *et al*., 2016; Ballerini *et al*., 2018) have applied 2D or 3D Frangi filter to detect vessel-like objects (Frangi *et al*., 1998). The Frangi vesselness is based on the eigenvalues of the Hessian matrix consisting of second-order derivatives of Gaussian filter responses. It describes the similarity of one structure to an ideal tube. Ballerini *et al*. (Ballerini *et al*., 2020; Ballerini *et al*., 2016; Ballerini *et al*., 2018) optimised the parameters in the Frangi vesselness function by applying an ordered logit model which took visual rating results into account. Given the various sizes of PVS, multi-scale vesselness was often calculated (Ballerini *et al*., 2020; Ballerini *et al*., 2016; Ballerini *et al*., 2018; Sepehrband *et al*., 2019).

Boespflug *et al*. (Boespflug *et al*., 2018) used the linearity and width of candidate clusters to remove FPs. Specifically, a line representing the eigenvector associated with the largest eigenvalue of the candidate cluster, was derived by using singular value decomposition. Norms were then fit from each voxel in the candidate cluster to the line. The correlation between the distance from each voxel to the cluster centre, and the distance from the point on the line defined by the voxel’s norm to the cluster centre, was considered as a measure of the cluster linearity.

The width of cluster was described as the summed distance of the two farthest points from the line and 1.7 (the corner-to-corner distance of a 1mm unit cube). A vessel-like, linear object should have high linearity (i.e., high correlation coefficients) and relatively small width. Similar approach was also applied to another study (Schwartz *et al*., 2019) to describe linearity, width and length of candidate clusters.

Descombes *et al*. (Descombes *et al*., 2004) applied template theory with a predefined prior of small tubular structures with a diameter of 1-3 mm and average length of 3-4 mm (max length, 15 mm). The model also took the interactions with surroundings into account.

#### 3.2.3.2 Classical machine learning

##### 3.2.3.2.1 Intensity features

Bag of words (BoW) is an approach to represent data, usually used in natural language processing. Gonzalez-Castro *et al*. (González-Castro *et al*., 2016) applied BoW to the task of PVS segmentation. Image patches around keypoints extracted from the basal ganglia region were first created through dense sampling. An intensity distribution descriptor was calculated to characterise the image patch. The k prototype vectors from the k-means clustering were considered as 'visual words', and a dictionary was created with these k visual words. All images in the dataset were used to build the dictionary of visual words. For each patch of an image in the dataset, the intensity distribution descriptor was calculated (patch-level descriptor). A visual word was considered as the representation of the patch when it had the shortest Euclidean distance from the patch-level descriptor. An image-level descriptor was obtained through considering the histogram of visual words representative of all patches. An SVM model was built with these image-level descriptors/features to predict image-level PVS visual rating scores (70% of all images as training set, and 30% as test set).

#### 3.2.3.2.2 Geometric and orientation features

In addition to the Frangi vesselness, similar to studies extracting CMB, other shape and orientation features have been derived from the Hessian matrix. These included the features derived from all 3 eigenvalues ($\frac{|\lambda_1|}{|\lambda_2|}, \frac{|\lambda_1|}{\sqrt{|\lambda_2 \lambda_3|}}, \sqrt{{\lambda_1}^2 + {\lambda_2}^2 + {\lambda_3}^2}$, where $\lambda_1 > \lambda_2 > \lambda_3$), and those resulting from the eigenvector $(x_1, y_1, z_1)$ corresponding to the largest eigenvalue $\lambda_1$ ($\arccos(\frac{|z_1|}{\sqrt{{x_1}^2 + {y_1}^2 + {z_1}^2}})$, $\arctan(\frac{y_1}{x_1})$) (Zhang *et al.*, 2016a, 2017).

In addition to the Hessian matrix-based features, Zhang *et al*. (Zhang *et al.*, 2016a, 2017) applied 2 other filters to describe the orientation PVS, namely the steerable filter and optimally oriented flux. The steerable filters are based on the fact that any arbitrarily oriented responses to the first- and second-order Gaussian derivative filters can be derived by linear combinations of some basis filter responses. In these two studies (Zhang *et al.*, 2016a, 2017), responses to 19 such basis filters (1 Gaussian filter, 9 oriented first-order Gaussian derivative filters, 9 oriented second-order Gaussian derivative filters) were used as features for the following RF classifier. Optimally oriented flux emphasises curvilinear structures by showing the amount of image gradient flowing in or out of a local sphere. Two sets of features were used: 1) features based on the first 2 eigenvalues of the 3×3 optimally oriented flux matrix ($\lambda_1, \lambda_2, \sqrt{\max(0, \lambda_1\lambda_2)}, \lambda_1 + \lambda_2$), and 2) features derived from the eigenvector corresponding to the largest eigenvalue (similar to the features extracted from Hessian matrix mentioned above).

Park *et al*. (Park *et al*., 2016) first identified candidates based on anatomical information and vesselness. Instead of specifying discriminative features, 3D Haar features were calculated in randomly defined image patches. Since the majority of informative features were located in the centre line, more features were extracted from the main direction of the vessel-like tube.

#### 3.2.3.3 Deep learning

Lian *et al*. (Lian *et al*., 2018a; Lian *et al*., 2018b) employed multi-channel inputs, including the original T2-weighted scan and the T2-weighted images enhanced for thin tubular structures, to provide both the information of regional intensities and the emphasis of tubular objects. In addition, the output from the initial run of the model was considered to provide contextual guidance for locating PVS, and therefore, recursively used as an additional input channel to refine the performance of the fully convolutional network. Sudre *et al*. (Sudre *et al*., 2019) used T1- and T2-weighted and T2-weighted FLAIR images as input, to detect and segment PVS and lacunes. This study considered the inconsistency between multiple raters, and used, instead of a hard classification, a soft probability label averaging among all raters as the gold standard.

#### 3.2.3.3.1 Weakly supervised learning

Since PVS are relatively small lesions and prevalent within human brains, visual rating is more practical compared to manual annotation when assessing PVS. The annotation of PVS was usually done in small samples. Therefore, a few studies had proposed strategies to make use of weakly labelled data (i.e., visual ratings) in designing deep learning models for PVS segmentation. Boutinaud *et al*. (Boutinaud *et al*., 2021) trained a convolutional autoencoder with visual rating data. The autoencoder had the same architecture as a U-Net model, which was used for the final segmentation, except for the absence of skip connections. Weights learnt from the autoencoder were transferred to the U-Net model for PVS segmentation. Dubost *et al*. (Dubost *et al*., 2020; Dubost *et al*., 2017) proposed a class activation mapping approach. During training a U-Net-like model to predict an image-level label (i.e., the number of PVS), a global pooling layer was added after the last convolutional layer and before a fully connected layer. In order to output the attention map, the global pooling layer was removed. The attention map was calculated as the linear combination of the feature maps (i.e. activation maps) derived from the last convolution layer (i.e., the layer before global pooling) using the weights of the following fully connected layer. Higher values in the attention map were more likely to be PVS.

### 3.2.3.4 Image enhancement for PVS segmentation

Studies have amplified the Haar transform coefficients corresponding to the desired signal (i.e., signal in PVS regions) to increase PVS visibility (Yang *et al*., 2021). However, this approach may amplify both signal and background noise when signal is not sufficiently discriminative. To solve such issue, Hou *et al*. (Hou *et al*., 2017) proposed a non-local strategy where the Haar transform was applied to a stack of neighbouring cubes of the reference cube. A nonlinear mapping function was then applied to amplify the signal and suppress the noise. This was followed by inverse Haar transform to generate enhanced neighbouring cubes. Enhanced neighbouring cubes at the same location corresponding to different reference cubes were finally averaged to form the entire enhanced MRI image. The residual noise was further reduced by applying a block-matching and 4D filtering algorithm which is also based on the non-local strategy. Jung *et al*. (Jung *et al*., 2019) proposed a deep 3D CNN to address the image enhancement task, and used the enhanced images according to Hou's approach (Hou *et al*., 2017) as ground truth. The proposed architecture included densely connected blocks with skip connections to reduce the number of parameters and mitigate the vanishing gradient problem in deep layers.

Sepehrband *et al*. (Sepehrband *et al*., 2019) applied an adaptive non-local mean filtering algorithm taking into account the surrounding voxels weighted by the Euclidean distance. The algorithm included a regularisation term to account for the Rician noise in MRI data. High-frequency noise was removed by applying a filtering patch with a radius of 1 voxel. The algorithm was applied to T1- and T2-weighted images separately, and the final enhanced PVS contrast was calculated as the enhanced T1-weighted image divided by the enhanced T2-weighted data.

## 3.3 Lacunes

### 3.3.1 Overview

Table 4 summarised studies of lacunes. We identified 2 published studies applying deep learning algorithms (Al-Masni *et al*., 2021; Ghafoorian *et al*., 2017), 2 using image processing techniques (Wang *et al*., 2012; Yokoyama *et al*., 2007), and a body of 6 studies by Uchiyama *et al*. combining image processing with machine learning methods (Uchiyama *et al*., 2015; Uchiyama *et al*., 2009; Uchiyama *et al*., 2012; Uchiyama *et al*., 2008; Uchiyama *et al*., 2007a, b), to the segmentation of lacunes and separate lacunes from other imaging biomarkers of cerebral small vessel diseases.

### 3.3.2 Materials

*Study samples* – While some studies did not specify details of included samples, others have used community-based cohorts (Wang *et al*., 2012), as well as patients with cerebral small vessel diseases, transient ischaemic attack, and/or ischaemic/haemorrhagic stroke (Ghafoorian *et al*., 2017). The median sample size in studies of lacunes was 132 (range, 100-1075), but due to the low prevalence of lacunes in general population, the number of lacunes was low (usually < 100).

*MRI modalities and preprocessing* – Lacunes appear hypointense on T1-weighted and T2-weighed FLAIR scans, and hyperintense on T2-weighted scans (Wardlaw *et al*., 2013). Therefore, these 3 MRI modalities or 2 out of the 3 modalities have been used in all included studies. Uchiyama *et al*. (Uchiyama *et al*., 2009) used magnetic resonance angiography (MRA) images in addition to T1-weighted and T2-weighted FLAIR scans to remove dilated perivascular spaces from lacune candidates. MRI preprocessing in segmenting lacunes usually included co-registration of different MRI modalities (Al-Masni *et al*., 2021; Ghafoorian *et al*., 2017; Wang *et al*., 2012), brain extraction (Ghafoorian *et al*., 2017; Wang *et al*., 2012), bias field correction (Ghafoorian *et al*., 2017; Wang *et al*., 2012), and intensity normalisation (Al-Masni *et al*., 2021; Ghafoorian *et al*., 2017).

*Defining true lacunes* – The ground truth labelling of lacunes were typically performed by experienced neuroradiologists and neurologists (Al-Masni *et al*., 2021; Uchiyama *et al*., 2007a). Ghafoorian *et al*. (Ghafoorian *et al*., 2017) applied the Standard for Reporting Vascular Changes on Neuroimaging (STRIVE), where lacunes were defined as round or ovoid, subcortical, fluid-filled cavities with 3-15 mm diameters (Wardlaw *et al*., 2013).

### 3.3.3 Algorithms

#### 3.3.3.1 Classical image processing and machine learning

##### 3.3.3.1.1 Intensity features

Wang *et al*. (Wang *et al*., 2012) applied intensity thresholding methods to segment lacunes adjacent to WMH and those located in subcortical structures. During candidate detection, since lacunes showed varying intensities in different stages (acute, subacute, chronic), multi-phase thresholding was used in a few studies to remove background (Uchiyama *et al*., 2007a; Yokoyama *et al*., 2007). On T2-weighted scans, it is difficult to segment lacunes adjacent to hyperintense regions (e.g., lateral ventricles) because of similar intensities. To address this issue, studies have also applied white top-hat transform to enhance lacunes with a user-defined structuring element (e.g., a square element (Uchiyama *et al*., 2007a), or circular elements with different radii (Yokoyama *et al*., 2007)).

##### 3.3.3.1.2 Geometric and location features

The largest and smallest eigenvalues of the Hessian matrix were used to describe nodular and linear objects (Uchiyama *et al*., 2015; Uchiyama *et al*., 2012; Uchiyama *et al*., 2007a). Nodular objects should have negative values for the second derivatives of Hessian eigenvalues in all directions. The second derivative of Hessian eigenvalues for linear objects should be close to zero in the direction along the long axis, and negative in the direction perpendicular to the long axis. These geometric features were calculated for multiple scales. The x- and y-coordinates were used to describe the location of lacune candidates (Uchiyama *et al*., 2015; Uchiyama *et al*., 2012; Uchiyama *et al*., 2007a). Template matching in eigenspace was also used to further reduce FPs (Uchiyama *et al*., 2015).

#### 3.3.3.2 Deep learning

Two recent studies have applied deep learning techniques. A fully convolutional network was used in the candidate detection phase to reduce the large amount of time consumed in the sliding window approach (Ghafoorian *et al*., 2017). For FP reduction, a 3D CNN was employed with multi-scale contextual information and hard-coded location features. Another study applied 3D multi-scale residual convolutional network with T1-weighted and T2-weighted FLAIR data as inputs (Al-Masni *et al*., 2021).

#### 3.3.3.3 Separating lacunes from PVS

Lacunes and PVS are both hypointense on T1-weighted and T2-weighted FLAIR scans, and hyperintense on T2-weighted images (Wardlaw *et al*., 2013). Therefore, a few efforts have been made to separate these two types of lesions. Uchiyama *et al*. (Uchiyama *et al*., 2009; Uchiyama *et al*., 2008) applied a 3-layer neural network with 6 features: 1) candidate locations described as x and y coordinates, 2) size of candidates, 3) degree of shape irregularity calculated as 1-C/L, where C is the length of circumference of the circle having the same area as the candidate, and L is the boundary length of the candidate, and 4) the intensity contrast between the candidate and its surrounding areas. Moreover, cerebral blood vessels were reconstructed from MRA, and superimposed onto T2-weighted scans, to aid the identification of PVS (Uchiyama *et al*., 2009).

# 4 Discussion

In a systematic review, we summarised 71 studies on computer-aided segmentations of three imaging biomarkers of cerebral small vessel diseases (CMB, PVS, lacunes). There is a trend of favouring artificial intelligence techniques (classical machine learning, and more recently, deep learning) in examining these biomarkers. Compared to CMB and PVS, lacunes are less well studied.

Since CMB, PVS and lacunes are small lesions, manually labelling them on MRI scans can be demanding. In particular, although PVSs are prevalent even in children and healthy young adults (Groeschel *et al*., 2006), they tend to be widely distributed in the brain with significantly varying numbers and locations between individuals, making manual labelling all PVSs almost impossible. As a result, studies had focused on certain highly vascularised areas (e.g., basal ganglia and centrum semiovale), and considered the PVS in these regions to be a valid representation of the severity of PVS in the whole brain. Despite being restricted to specific areas, many studies still only had access to visual rating data (e.g., the number of PVS, and categories of PVS severity according to the number). This raised the issue of weakly labelled data for PVS.

On the other hand, CMB and lacunes are less common to a significant degree. In the general population, CMB was present in 7-13% of individuals aged 40-69 years (Lu et al., 2021a; Poels et al., 2010). The majority of these individuals had a single CMB (71.9-80.6% of individuals showing CMB (Lu et al., 2021a; Poels et al., 2010)). Individuals over 80 years of age (35.7% prevalence (Poels *et al*., 2010)) and stroke patients (29.4%, (Ibrahim *et al*., 2019)) had a relatively higher prevalence. Lacunes were reported to be found in 8-31% of the healthy older individuals (Das *et al*., 2019). Similar to CMB, most individuals (66% of those with lacunes) had a single lacune. The low prevalence of CMB and lacunes has limited the size of training data.

A few novel approaches have been applied to accommodate the weakly labelled data and limited data sample sizes. In image processing tasks using CNN, the early layers typically learn simple features such as edges and contrasts. Additional layers then follow to learn more abstract features, with the last layers trying to map from learnt features to the task-dependent output (e.g., number of lesions). Since the features learnt in early layers are not expected to be specific to the task, it is possible to learn these features from other large datasets with similar semantics. The parameters in the early layers can then be frozen, and later layers can be adjusted to address the particular task using the gold-standard labels. This representation learning technique is usually referred to as transfer learning. It significantly reduces the required sample sizes for training a CNN model. This has been applied in some studies to address the issue of small training sample sizes (Afzal *et al*., 2022; Hong *et al*., 2019; Li *et al*., 2021). Some of these studies used models pre-trained on the ImageNet dataset (Afzal *et al*., 2022; Li *et al*., 2021), a large-scale dataset of natural images. However, natural images differ in many ways from MRI scans. Natural images are often colour images with RGB channels with a few salient objects, whereas MRI scans are grayscale images indicating magnetic features of brain tissue with less distinction between foreground and background. Moreover, MRI scans of the brain are 3D images with roughly consistent locations and contrasts for brain structures, though individual differences exist. Therefore, models pre-trained on large MRI datasets with image-level labels (e.g., age, sex) can be promising solutions (Nicola *et al*., 2021). Moreover, since there have been established pipelines to automatically segment WMH, another biomarker of CSVD, future studies may consider pre-training on the severity of WMH.

For the issue of image-level weak labels, some studies modified the U-Net architecture for segmentation to first resolve a task of predicting these weak labels (i.e., a regression task). The learnt weights were then transferred to the final segmentation model (Boutinaud *et al*., 2021). Alternatively, the global pooling layer can be removed to generate activation maps indicating the probability of lesions (Dubost *et al*., 2020). Future studies may also consider having these results checked and modified by domain experts to make gold-standard labels for segmentation. This will save much time compared to labelling lesions from scratch.

There are three dimensions regarding training data that need to be considered: size, diversity, and accuracy. Although a few techniques can be applied to reduce the required amount of data for training, a decent sample size is necessary to tune the network parameters for a specific task, and to avoid overfitting issues. However, including inaccurately labelled data will compromise the model performance. In addition, a large number of similar cases contribute less than a good variety of data. The diversity of data can come from different sample types (healthy vs. various diseases) and scanners/scanning parameters. On most occasions, these three factors cannot be all satisfied. Labelling small lesions, like CMB, PVS and lacunes, at the voxel level is time-consuming and subjective. Usually, multiple raters are assigned with the same set of data for labelling. Results are then compared, and for inconsistent labels, a consensus is reached by the majority vote, or being checked by a more experienced clinician. As a result, individual studies are unlikely to obtain sufficient high-accuracy voxel-level labels for training. Collaborations are a promising approach in this regard, as it not only pools small data samples with labels, but also increases data diversity in the meantime. Public resources identified during the systemic review have been summarised in Table 5. Data augmentation is another approach. Available studies have applied basic translations, rotations and mirroring to generate more data (Rashid *et al.*, 2021). Introducing artefacts such as bias field and noise to the images may be helpful for the model to generalise to different samples/scanners (Billot *et al.*, 2020). In addition, techniques such as uncertainty modelling (Karimi et al., 2020) and domain adaption (Guan and Liu, 2022) can be incorporated to address the issues due to inaccurate or subjective labels and excessive data heterogeneity due to the data acquisition process. Such methods have been developed for biomedical image analysis in general, but not particularly for the domain of CSVD lesions.

The MRI modalities to input into models are also worth considering carefully. Although there have been consensus MRI modalities for each CSVD lesion, a small number of studies included additional modalities (e.g., QSM (Rashid *et al*., 2021) and PD (Ghafaryasl *et al*., 2012)), that are not commonly available, to enhance the performance. This may not be optimal when considering the generalisability of the model. Some techniques, such as knowledge distillation (Guan *et al*., 2021), could be considered to resolve this issue.

The performance of models would be more comparable if there is a reference dataset with a good diversity of data where the performance metrics can be calculated and compared. To this end, the 'Where is VLADO – Vascular Lesions Detection Challenge' (https://valdo.grand-challenge.org/Description/) is a good initiative. Training data with labels are provided, and a hidden dataset is used to compare performance. The selection of performance metrics should consider the clinical importance of the measures. For example, studies showed stronger associations of the volume, length, width and size of PVS with WMH burden (Ballerini *et al*., 2020), a well-established biomarker for CSVD, compared to the number of PVS. This emphasised the importance of segmenting PVS, and therefore voxel-wise performance metrics should be used. On the other hand, due to the low prevalence, studies of CMB (Lu *et al*., 2021a) and lacunes (Ghaznawi *et al*., 2019) usually examined the number or presence of lesions. As a result, cluster- or image-level performance metrics may be more clinically relevant. Moreover, we propose that clinical validations (correlating with demographic and/or clinical measures) should supplement image processing performance metrics when assessing the models. Most available studies validated the algorithms through calculating performance metrics (e.g., sensitivity, specificity, etc.) with few investigating the relationship with clinical findings (Ballerini *et al*., 2020; Ballerini *et al*., 2016; Ballerini *et al*., 2018). Although our ultimate goal is to accurately segment CSVD lesions, missing a small proportion of lesions or lesion voxels usually do not significantly alter the correlation with clinical measures when studying large cohorts. Therefore, as the first step, it makes sense to aim at models with reasonable performance measurements which produce meaningful clinical associations in large datasets (e.g., UK Biobank (Miller et al., 2016), OASIS (Marcus et al., 2010)).

In conclusion, advancements in artificial intelligence have produced promising results in the segmentation of CSVD lesions. Future studies could consider pooling data from different sources, techniques for weak supervision and transfer learning, and assessing the model performance with both image processing metrics and clinical correlations.

# 5 Tables

Table 1. Definition of cerebral microbleed, perivascular space, and lacune of presumed vascular origin on MRI scans

| Lesion | Usual MRI modalities | Intensity | Shape | Size |
|---|---|---|---|---|
| Cerebral microbleed | T2*w GRE, SWI | Dark | Round/ovoid | Diameter = 2-5 mm (can be up to 10 mm). Blooming with longer TE. |
| Perivascular space | T1w, T2w, T2w-FLAIR | Dark on T1w and T2w-FLAIR. Bright on T2w. | Round/ovoid in the plane perpendicular to the course of vessel. Linear in the plane parallel to vessel. | Diameter ≤ 3mm |
| Lacune of presumed vascular origin | T1w, T2w, T2w-FLAIR | Dark on T1w and T2w-FLAIR. Bright on T2w. Usually have bright rim on T2w-FLAIR. | Round/ovoid | Diameter = 3-15 mm |

T2*w GRE, T2*-weighted gradient recalled echo sequence; SWI, susceptibility-weighted imaging; TE, echo time; T1w, T1-weighted imaging; T2w, T2-weighted imaging; T2w-FLAIR, T2-weighted fluid-attenuated inverse recovery sequence.

Table 2. Summary of CMB studies.

| Study | method | sample type | Total N of ppts / CMBs | MRI modalities used for segmentation | Task | MRI preproc | Algorithm | Performance |
|---|---|---|---|---|---|---|---|---|
| (Chen et al., 2015) | DL | transient ischemic attack | 20/117 | SWI | S | intensity normalisation | intensity thresholding + CNN + SVM | Sensitivity=0.8913; Precision=0.5616; Avg FPs = 6.4 |
| (Zhang et al., 2016b) | DL | CADASIL; healthy controls | 10 | SWI | S | | DNN with sparse autoencoder | Sensitivity = 93.20+/-1.37%; Specificity = 93.25+/-1.38%; Accuracy = 93.22+/-1.37% |
| (Dou et al., 2016) | DL | stroke; normal ageing | 320/1149 | SWI | S | intensity normalisation | 3D FCN + CNN | Sensitivity = 93.16%; an average of 2.74 false positives per subject; precision = 44.31% |
| (Wang et al., 2017) | DL | CADASIL | 10 | SWI | S | | CNN with rank-based average pooling | sensitivity = 96.94%; specificity = 97.18%; accuracy = 97.18% |
| (Lu et al., 2017) | DL | | 64 | SWI | S | | CNN | sensitivity = 97.29%; specificity = 92.23%; accuracy = 96.05% |
| (Zhang et al., 2018a) | DL | CADASIL; healthy controls | 20 | SWI | S | | CNN | sensitivity = 93.05%; specificity = 93.06%; accuracy = 93.06% |
| (Zhang et al., 2018b) | DL | CADASIL; healthy controls | 20 | SWI | S | | DNN (one input layer, four sparse autoencoder layers, one softmax layer, and one output layer) | |
| (Hong et al., 2019) | DL | CADASIL | 10 | SWI | S | | CNN (ResNet50) with transfer learning | sensitivity = 95.71+/-1.044%; specificity = 99.21+/-0.076%; accuracy = 97.46+/-0.524% |
| (Liu et al., 2019) | DL | hemodialysis; TBI; stroke; healthy controls | 220/1641 | SWI | S | bias field correction; warp to standard space | 3D FRST + ResNet (SWI and high-pass filtered phase image) | sensitivity = 95.8%; precision = 70.9%; 1.6 FPs per case |
| (Wang et al., 2019) | DL | CADASIL; healthy controls | 20 | | S | | DenseNet; transfer learning (freeze earlier layers and train later layers only) | sensitivity = 97.78%; specificity = 97.64%; accuracy = 97.71%; precision = 97.65% |
| (Chen et al., 2019) | DL | individuals who had giomas, underwent radiation therapy, with confirmed radiation-induced CMBs | 73 | SWI | S | | 3D ResNet | precision = 71.9%; 94.7% of true CMBs were correctly identified |
| (Hong et al., 2020) | DL | CADASIL | 10 | SWI | S | | CNN | sensitivity = 99.74+/-0.13%; specificity = 96.89+/-0.26%; accuracy = 98.32+/-0.15% |
| (Al-Masni et al., 2020) | DL | | 179/760 | SWI | S | brain extraction; data augmentation | YOLO detection + 3D CNN | High resolution data: Avg num of FPs per subject = 1.42; Low resolution data: Avg num of FPs per subject = 1.89 |
| (Rashid et al., 2021) | DL | community | 24 / 4 had no CMB; 13 had 1 or 2; 6 had 3-8; 1 had >100 | T2w, SWI, QSM | S | bias field correction; coregistration; brain extraction; intensity normalisation; data augmentation. | U-Net with a higher number of resolution layers and padded convolutions. | sensitivity = 0.84-0.88; precision = 0.40-0.59 |
| (Lu et al., 2021b) | DL | | | SWI | S | | CNN + extreme learning machine | sensitivity = 94.53%; specificity = 96.10%; accuracy = 95.25% |
| (Myung et al., 2021) | DL | | 186/1133 | GRE | S | skull stripping; data augmentation | YOLO detection +/- CSF filtering | sensitivity = 59.69-80.96%; precision = 60.98-79.75%; FPs per subject = 2.15-6.57 |
| (Li et al., 2021) | DL | | 58 | SWI | S | bias field correction | CNN with transfer learning (SSD-512/VGG pre-trained on ImageNet dataset) + feature enhancement | sensitivity=90%; precision=79.7% |
| (Afzal et al., 2022) | DL | | 20 | SWI | S | skull stripping | CNN with transfer learning (AlexNet and ResNet50 pre-trained on ImageNet dataset) | AlexNet: accuracy=97.26%, 1.8% FP rate; ResNet50: accuracy=97.89%, 1.1% FP rate |
| (Barnes et al., 2011) | ML | MCI; early AD | 126 CMBs | SWI | D | brain extraction; interpolation | intensity thresholding + SVM (covariance matrix shape, intensity, and size features) | Sensitivity = 81.7%; Specificity = 95.9% |
| (Ghafaryasl et al., 2012) | ML | | 237/631 | GRE, PD | S | bias field correction, coregistration, brain extraction | intensity and size thresholding + LDC/QDC/Parzen (size, shape, intensity features) + LDC/QDC/Parzen/linear SVM (Hessian-based shape, intensity on PD) | Sensitivity = 91%; On avg, 4.1 FPs per ppt |
| (Fazlollahi et al., 2014) | ML | AD; MCI; healthy | 41/104 | SWI | S | tissue segmentation; brain extraction; bias field correction; intensity normalisation; inverting image contrast so that CMBs and vessels appear hyperintense; applying gradient based anisotropic diffusion and adaptive histogram equalisation | multi-scale Laplacian of Gaussian spherical/semi-spherical object detection + multi-layer RF (Radon-based shape features) | sensitivity = 92.04%; on average 6.7 FPs per true CMB and 16.84 false CMBs per subject |
| (Fazlollahi et al., 2015) | ML | AD; MCI; healthy | 66/231 | SWI | S | same as (Fazlollahi 2014) | multi-scale Laplacian of Gaussian spherical/semi-spherical object detection + multi- | Sensitivity of 87% and avg false detection rate of 27.1 CMBs per ppt on the 'possible and definite' set; Sensitivity of |
| | | | | | | | layer RF (Radon- and Hessian-based shape features) | 93% and avg false detection rate of 10.0 CMBs per ppt on the 'definite' set. |
| (Roy et al., 2015) | ML | mild to severe traumatic brain injury | 27 | SWI | S | SWI phase enhancement; brain extraction; bias field correction; registering T1w to SWI to obtain WM mask | RST + RF (intensity, RST shape features) + size thresholding | Sensitivity = 85.6%; Specificity = 99.5% |
| (Dou et al., 2015) | ML | | 44/615 | SWI | S | intensity normalisation | intensity and size thresholding + RF (intensity features) + 2-layer ISA for feature extraction + SVM | Sensitivity = 89.44%; Avg of 7.7 and 0.9 FPs per subject and per CMB, respectively |
| (van den Heuvel et al., 2015; van den Heuvel et al., 2016) | ML | moderate to severe traumatic brain injury; healthy controls | 33-51 | T1w, SWI | S | tissue segmentation; coregistration; bias field correction; intensity normalisation | intensity thresholding + voxel-level RF (intensity,shape) + region growing + region-level RF (intensity, shape, size) + manual correction | sensitivity = 89.1±0.8% / 93.2±1.0% with on average 25.9±0.8 / 12.9±0.8 FPs per TBI patient |
| (Seghier et al., 2011) | IP | unselected, consecutive patients to Stroke Service | 30 | GRE | S | | consider CMB as an extra class in unified normalisation-segmentation in SPM; a flat prior for CMB in the first pass, then modify subject-specific prior; 2nd pass with refined priors | Kappa = 0.43 (0.65 after manual correction); ICC = 0.71 (0.87 after manual correction) |
| (Kuijf et al., 2011) | IP | individuals with CMB, no other brain pathology | 2 | dual-echo SWI | S | Minimal intensity projection post-processing for both echos simultaneously; Processed images with a slab thickness of 2mm (no overlap), and images with a slab thickness of 4mm and a -2mm gap between the slices (2 mm overlap) were used for both scans. | 3D RST | |
| (Kuijf et al., 2012) | IP | individuals with CMB | 18/54 | dual-echo SWI | S | brain masking containing GM and WM; intensity normalisation | 3D RST | Sensitivity = 71.2% |
| (Kuijf et al., 2013) | IP | consecutive patients to memory clinic | 72 | GRE | S | | 3D RST | sensitivity = 65-84% depending on settings |
| (Bian et al., 2013) | IP | individuals who had giomas, underwent radiation therapy, with confirmed radiation-induced CMBs | 15 | SWI | S | brain extraction, intensity normalisation | 2D FRST + intensity thresholding on FRST map + vessel mask screening (using FRST orientation projection map) + 3D region growing + 2D geometric feature examination (area, circularity) | sensitivity for 'definite' = 95.4%; sensitivity for 'possible' = 77.5%; sensitivity for 'total' = 86.5% |
| (Tajudin et al., 2017a; Tajudin et al., 2017b) | IP | | | | S | | CLAHE to improve contrast + Bottom-Hat filtering + K-means + watershed + active contour segmentation | |
| (Morrison et al., 2018) | IP | individuals who had giomas, underwent radiation therapy, with confirmed radiation-induced CMBs | 15/248 | GRE | S | brain extraction, intensity normalisation | (Bian et al., 2013) + volume segmentation using local intensity thresholding + 2D geometric feature examination (circularity, centroid-to-seed point distance) + size thresholding + manual judgement | Sensitivity = 86.7% (215/248 CMBs detected) |
| (Liu et al., 2020b) | IP | | 20 | | S | | dual domain (gradient and voxel) distribution + distribution of Fourier coefficients | sensitivity=85.2%; precision=3.2% |
| (Chesebro et al., 2021) | IP | community | 78/64 | SWI, GRE | S | brain extraction, coregistration, applying lobar masks, upsampling by factor of 3 | Sobel filter + Canny edge detection + circular Hough transform + entropy of neighbourhood + Frangi filter-based analysis + manual correction | sensitivity = 95% |

DL, deep learning; ML, machine learning; IP, image processing; CADASIL, cerebral autosomal dominant arteriopathy with subcortical infarcts and leukoencephalopathy; MCI, mild cognitive impairment; AD, Alzheimer's Disease; SWI, susceptibility-weighted imaging; QSM, quantitative susceptibility mapping; GRE, gradient echo sequence; PD, proton density-weighted imaging; S, segmentation; D, detection; CNN, convolutional neural network; SVM, support vector machine; FCN, fully convolutional network; RST, radial symmetry transform; FRST, fast RST; YOLO, You Only Look Once; CSF, cerebrospinal fluid; SSD-512/VGG, VGG-based Single Shot Multi-box Detector network; LDC, linear discriminant classifier; QDC, quadratic discriminant classifier; RF, random forest; ISA, independent subspace analysis.

Table 3. Summary of PVS studies

| Study | Method | Sample type | Total N of ppts | MRI modalities | Brain region/s | Task | MRI preproc | Algorithm | Performance |
|---|---|---|---|---|---|---|---|---|---|
| (Descombes et al., 2004) | IP | | 37 | T1w | whole brain | S | | A model considering intensity, location, geometric properties. Reversible Jump Markov Chain Monte Carlo algorithm to optimise the model. | Type I error of 13.0% (FPs) and type II error of 2.7% (FNs); ICC with visual = 0.87 |
| (Wang et al., 2016) | IP | mild stroke patients | 100 | T2w | BG | S | coregistration different modalities, brain extraction | adaptive intensity thresholding | BG PVS count associated with WMH and global atrophy |
| (Ballerini et al., 2016) | IP | community | 24 | T2w | CS | S | brain extraction, tissue segmentation, locating CS, reslicing | 3D Frangi filter. Ordered Logit Model (OLM) to optimise Frangi filter parameters. | correlated with neuroradiological assessments |
| (Hou et al., 2017) | IP + ML | healthy volunteers (25-37 yo) | 17 | T2w | whole brain | E | | Haar transform-based non-local line singularity representation | Dice = 0.75; sensitivity = 0.77; PPV = 0.73 |
| (Boespflug et al., 2018) | IP | community-based non-demented | 14 * 2 timepoints | T1w, FLAIR, T2w, PD | whole brain | S | tissue segmentation, bias field correction, registering other modalities to T1w | based on 1) relative normalised intensity in WM, ventricle and cortex on T1w, FLAIR, T2w, and PD data. 2) Morphologic (width, volume, linearity) characterization of each cluster. | correlations with manual counts =0.54-0.69 |
| (Ballerini et al., 2018) | IP | dementia, stroke | 68 | T1w, T2w | CS | S | tissue segmentation, CSF+GM mask to reduce FPs, hole filling in WM, identifying CS ROI, reslicing. | same as (Ballerini et al., 2016): 3D Frangi filter with OLM optimisaion. | correlated with neuroradiological assessments |
| (Sepehrband et al., 2019) | IP | HCP, healthy young | 100 | T1w, T2w | whole brain | E/S | correction for gradient nonlinearity, readout, bias field; AC-PC alignment; warp to MNI; tissue segmentation. | adaptive non-local mean filtering applied to T1w and T2w, enhanced contrast calculated as T1w/T2w, multi-scale Frangi vesselness filters | better PVS visibility, more PVS identified by rater, PVS count on enhanced and original image were correlated, high inter-rater and test-retest reliability using enhanced image, high |
| (Schwartz et al., 2019) | IP | community, ADNI | 44 | T1w, FLAIR | CS | S | tissue segmentation, bias field correction | local intensity inhomogeneity search on white matter-masked T1-weighted data, constrained by FLAIR hyperintensities, as well as width, volume, and linearity measurements | correlation with visual counts = 0.72; test-retest (r=.87 (single slice), .77 (whole brain)); PPV = 77.5-87.5% |
| (Ballerini et al., 2020) | IP | community (mean age, 72.6) | 700 | T2w | CS, deep corona radiata | S | generating individual CS mask through warping a CS mask in | same as (Ballerini et al., 2016): 3D Frangi filter with OLM optimisaion. | correlation with visual = 0.47-0.61; associated with hypertension, stroke, WMH. |
| | | | | | | | standard space, reslicing | | |
| (Liu et al., 2020) | IP | children with idiopathic generalised epilepsy/controls | 62 | T2w | WM above ventricles | S | skull stripping, tissue segmentation | 2D Frangi filter | |
| (González-Castro et al., 2016) | ML | stroke patients | 264 | T2w | BG | S | brain extraction, tissue segmentation, subcortical structure extraction, detecting BG ROI, CLAHE, num of bright blobs to determine best slice for PVS. | Bag of visual words (BoW) with Scale Invariant Feature Transform (SIFT). | accuracy = 82.34% |
| (Park et al., 2016) | ML | healthy volunteers (25-37 yo) | 17 | T2w | whole brain | S | define candidates by anatomy and vesselness | randomised Haar features, sequential learning. | Dice = 0.63±0.05; sensitivity = 0.59±0.08; PPV = 0.68±0.06 |
| (González-Castro et al., 2017) | ML | stroke patients | 264 | T2w | BG | S | same as (González-Castro et al., 2016) | compared 3 descriptors: 1) Wavelet transform coefficients, 2) local binary patterns, and 3) BoW with SIFT. | accuracy (SVM+BoW+SIFT) = 81.16% - best among 3; high agreement with visual ratings. |
| (Zhang et al., 2016a, 2017) | ML | | 19 | T2w | whole brain | S | | entropy-based sampling to extract informative samples, responses to 3 filters (steerable filters, | Dice = 0.66 |
| | | | | | | | | optimally oriented flux, Frangi filter) as features. | |
| (Dubost et al., 2017) | DL | community | 1642 | PD | BG | D | Warp to MNI, ROI masks from FreeSurfer, intensity normalisation | Compare to (Dubost et al., 2020): deeper architecture, no residual connections. | sensitivity = 62%; avg 1.5 FPs per image |
| (Lian et al., 2018a, 2018b) | DL | healthy young to middle age | 17-20 | T2w | whole brain | S | | multi-channel FCN; 2 inputs: original T2w, and T2w after denoising and enhencing for tubular structures (BM4D, Haar transform-based line sigularity representation); auto-context knowledge (recursive use of PVS probability maps from initial run). | Dice = 0.78±0.07 / 0.77±0.06; sensitivity = 0.74±0.12 / 0.74±0.13; PPV = 0.83±0.05 / 0.83±0.06 |
| (Sudre et al., 2019) | DL | | 16 | T1w, T2w, FLAIR | whole brain | D/S | bias field correction, skull stripping, z-tranform to WM region statistics | 2D RCNN for multi-class multi-instance simultaneous detection and segmentation of PVS and lacunes. | sensitivity = 72.7%; median overlap of 59% between detected bounding box and ground truth when all raters agreed, and an overlap of 30% for more uncertain cases (>= 1 rater considering non-lesion). |
| (Dubost et al., 2019b) | DL | community | 2115 | T2w | midbrain, hippocampi, BG, CS | R | Bias field correction, ROI mask from FreeSurfer, individual mask adjustment, intensity normalisation. | Compare to (Dubost et al., 2019a): a) simpler and lighter, b) including skip connections, and c) global pooling instead of 2 fully connected layers. | ICC against visual = 0.75-0.88; ICC of test-retest = 0.82-0.93; similar associations with 20 PVS determinants to visual scores. |
| (Dubost et al., 2019a) | DL | community | 2017 | PD | BG | R | Warp to MNI, ROI masks from FreeSurfer, intensity normalisation | regression CNN to predict num of PVS, with more convolutional than pooling layers and no final activation. | ICC against visual = 0.74; ICC of test-retest = 0.93; correlated with age similar to visual. |
| (Jung et al., 2019) | DL | healthy volunteers (25-37 yo) | 17 | T2w | whole brain | E | | 3D CNN with dense skip connections to alleviate the gradient vanishing problem. | peak SNR = 47.98±2.66 (PVS), 44.59±2.23 (WM); structural similarity with gold standard = 0.975±0.01 (PVS), 0.98±0.01 (WM). |
| (Dubost et al., 2020) | DL | | 2202 | T2w | midbrain, hippocampi, BG, CS | D | FreeSurfer segmentation of ROI, intensity normalisation | Train a regression CNN with total num of PVS. During inference, attention maps of PVS were calculated as linear combination of feature maps of the layer before global pooling layer using weights from the following fully connected layer. | sensitivity = 62.1±8.7%; avg 2.33±1.71 FPs per scan; avg 2.44±2.01 FNs per scan |
| (Yang et al., 2021) | DL | young healthy/dementia/MCI/subjective memory impairment | 96 | T2w | BG | R | intensity normalisation, identify BG slices, Haar to enhance visibility. | CNN | Accuracy = 87.7% (image-level), 80.3% (subject-level) |
| (Boutinaud et al., 2021) | DL | university students | 1832 | T1w | BG, deep WM | S | tissue segmentation, skull stripping, intensity normalisation, generating BG mask | 3D convolutional autoencoder trained on visual rating data, transferring learnt weights to U-Net | Dice (voxel-level) = 0.51 (DWM), 0.66 (BG); Dice (cluster-level) = 0.64 (DWM), 0.71 (BG); Dice > 0.9 for large PVS; R^2 > 0.38 for DWM, and 0.02 for BG against visual rating. |

DL, deep learning; ML, machine learning; IP, image processing; HCP, human connectome project; ADNI, Alzheimer's Disease Neuroimaging Initiative; BG, basal ganglia; CS, centrum semiovale; S, segmentation; E, image enhancement; D, detection; R, regression; CSF, cerebrospinal fluid; GM, grey matter; WM, white matter; FP, false positive; ROI, region of interest; CNN, convolutional neural network; ICC, intraclass correlation coefficient.

Table 4. Summary of lacune studies

| Study | Method | Sample type | Total N of ppts/lacunes | MRI modalities | Task | MRI preproc | Algorithm | Performance |
|---|---|---|---|---|---|---|---|---|
| (Yokoyama et al., 2007) | IP | | 100 | T1w, T2w | S | | multiple-phase binarization for isolated lacunes; white top-hat transform with circular structuring elements for lacunes adjacent for hyperintense structures; FP reduction using features on T1w | sensitivity = 90.1%; specificity = 30.0%; 1.7 FPs per image |
| (Uchiyama et al., 2007b) | IP + ML | | 132 | T1w, T2w | S | | same as (Uchiyama et al., 2007a), but replacing SVM with a 3-layer neural network followed by modular classifiers to reduce 3 types of FPs | sensitivity = 96.80%; 0.30 FP per slice |
| (Uchiyama et al., 2012; Uchiyama et al., 2007a) | IP + ML | | 132 | T1w, T2w | S | brain extraction with region growing | candidate identificatin using white top-hat transform and multi-phase binarisation; FP reduction considering location, intensity and shape. | sensitivity = 96.80%; 0.76 FP per slice |
| (Uchiyama et al., 2009; Uchiyama et al., 2008) | IP + ML | | 109/89 | T1w, T2w, (MRA in 2009 study) | S | | white top-hat transform; intensity thresholding on transformed images; 3-layer neural network to separate lacunes from PVS (+ MRA to identify and separate PVS from lacunes) | sensitivity = 93.3% (83/89); specificity = 75.0% (15/20) |
| (Wang et al., 2012) | IP | community | 272/62 | T1w, T2w, FLAIR | S | bias field correction, coregistration, brain extraction, tissue segmentation, WMH extraction | lacunes near WMH: dilation + intensity contrasts; lacunes in subcortical structures: intensity thresholding. | sensitivity = 80.6% (detected 50/62 lacunes) |
| (Uchiyama et al., 2015) | IP + ML | | 132 | T1w, T2w | S | | same as (Uchiyama et al., 2007a). Add template matching in eigenspace for FP reductions | sensitivity = 96.80%; 0.47 FP per slice |
| (Ghafoorian et al., 2017) | DL | non-demented older individuals, young stroke patients | 1075 | T1w, FLAIR | S | coregistration, warp to standard space, bias field correction, intensity normalisation | FCN + 3D CNN (incld. multi-scale contextual info, and 7 explicit location features) | sensitivity = 97.4%; 0.13 FPs per slice |
| (Al-Masni et al., 2021) | DL | | 288/696 | T1w, FLAIR | S | coregistration, intensity normalisation | manual identification of candidates, 3D ResNet | sensitivity = 96.41%; specificity = 90.92%; accuracy = 93.67%; precision = 91.40%; avg FPs per ppt = 1.32 |

| | |
|---|---|
| (Sudre et al., 2019) | (Sudre et al., 2019) segmented both lacunes and PVS. Refer to Table 3 for more details. |

DL, deep learning; ML, machine learning; IP, image processing; S, segmentation, WMH, white matter hyperintensity; FP, false positive; SVM, support vector machine; MRA, magnetic resonance angiography; PVS, perivascular spaces; FCN, fully convolutional network; CNN, convolutional neural network.

Table 5. Publicly available code/datasets

| Link | Description | Reference |
|---|---|---|
| https://github.com/Yonsei-MILab/Cerebral-Microbleeds-Detection | code for CMB segmentation | (Al-Masni et al., 2020) |
| https://github.com/Yonsei-MILab/Lacunes-Identification | code for lacune segmentation | (Al-Masni et al., 2021) |
| https://appsrv.cse.cuhk.edu.hk/~qdou/cmb-3dcnn/cmb-3dcnn.html | code and labelled data for CMB segmentation | (Dou et al., 2016) |
| https://github.com/hjkuijf/MixLacune | code for lacune segmentation | - |
| https:// valdo.grand-challenge.org/ | "Where is VALDO" challenge, including segmenting PVS, CMB and lacunes, with example training data | - |

# 6 Figures

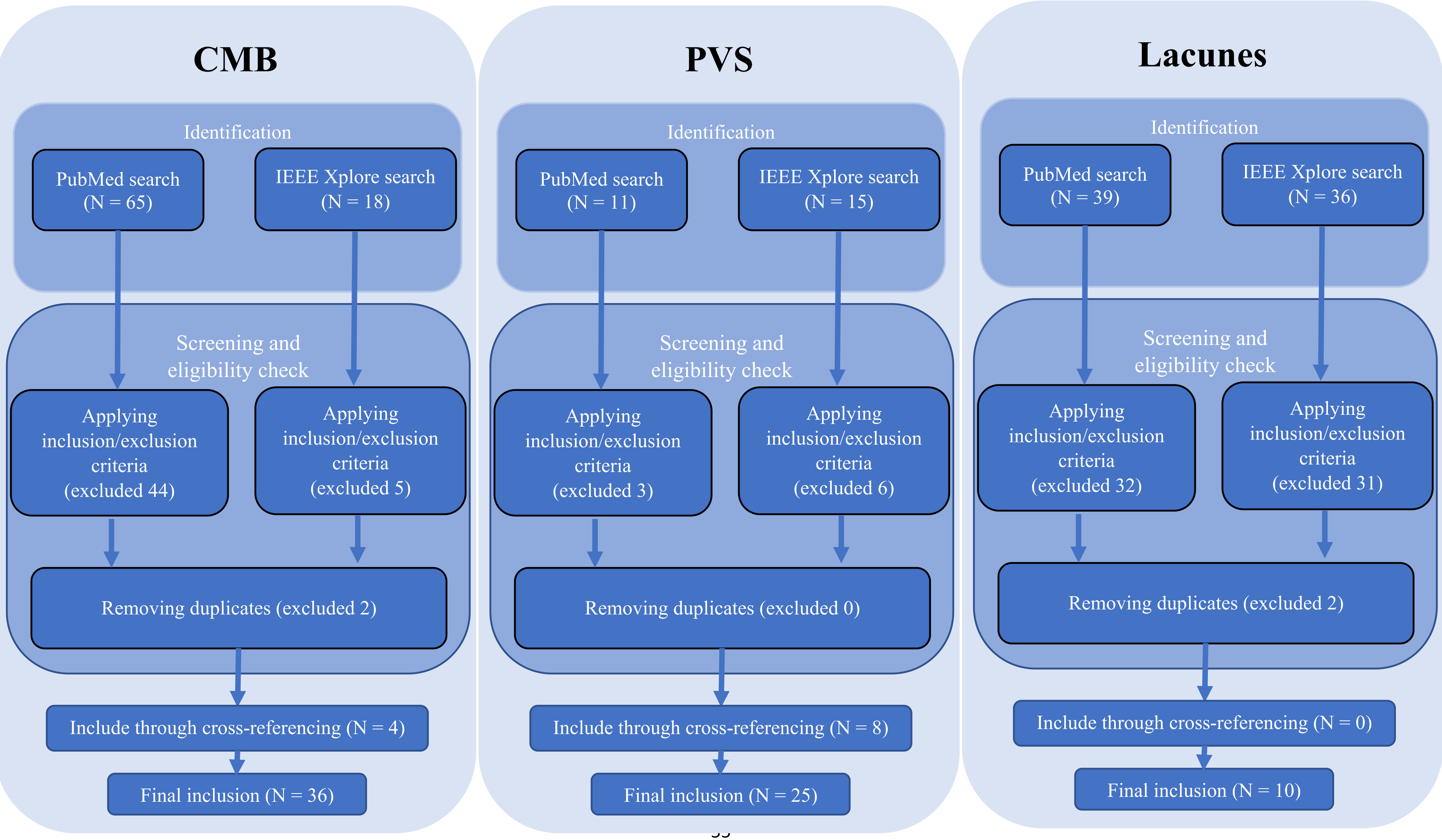
CMB
Identification
PubMed search (N = 65)
IEEE Xplore search (N = 18)
Screening and eligibility check
Applying inclusion/exclusion criteria (excluded 44)
Applying inclusion/exclusion criteria (excluded 5)
Removing duplicates (excluded 2)
Include through cross-referencing (N = 4)
Final inclusion (N = 36)
PVS
Identification
PubMed search (N = 11)
IEEE Xplore search (N = 15)
Screening and eligibility check
Applying inclusion/exclusion criteria (excluded 3)
Applying inclusion/exclusion criteria (excluded 6)
Removing duplicates (excluded 0)
Include through cross-referencing (N = 8)
Final inclusion (N = 25)
Lacunes
Identification
PubMed search (N = 39)
IEEE Xplore search (N = 36)
Screening and eligibility check
Applying inclusion/exclusion criteria (excluded 32)
Applying inclusion/exclusion criteria (excluded 31)
Removing duplicates (excluded 2)
Include through cross-referencing (N = 0)
Final inclusion (N = 10)

Figure 1. Literature search process.

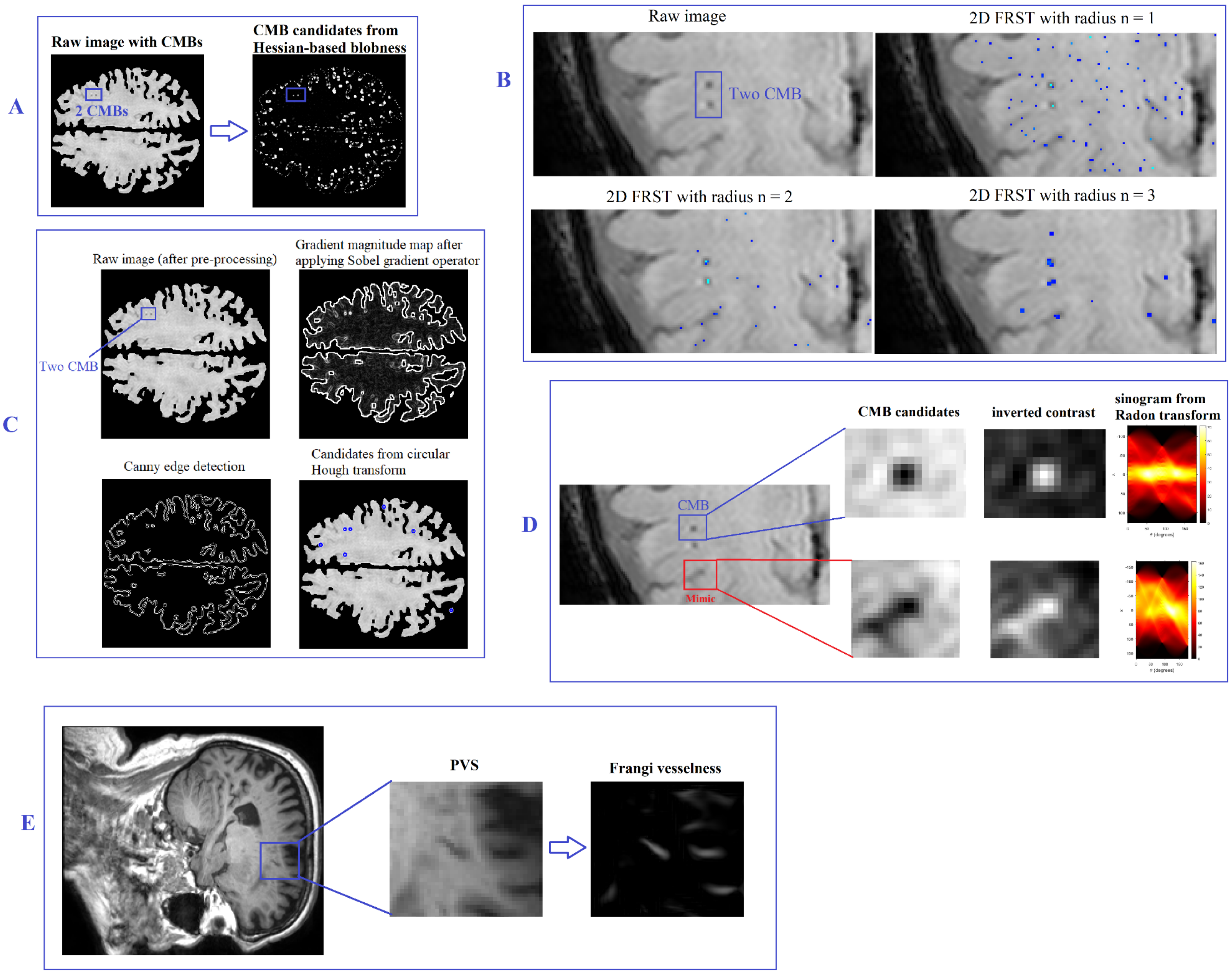

A
Raw image with CMBs
2 CMBs
CMB candidates from Hessian-based blobness
B
Raw image
Two CMB
2D FRST with radius n = 1
2D FRST with radius n = 2
2D FRST with radius n = 3
C
Raw image (after pre-processing)
Two CMB
Gradient magnitude map after applying Sobel gradient operator
Canny edge detection
Candidates from circular Hough transform
D
CMB
Mimic
CMB candidates
inverted contrast
sinogram from Radon transform
E
PVS
Frangi vesselness

Figure 2. Illustration of some classical image processing techniques. (A) Hessian matrix-based blobness; (B) 2D fast radial symmetry transform; (C) Sobel gradient map, Canny edge detection, and circular Hough transform; (D) Radon transform; (E) Frangi vesselness filter.

# 7 References

Afzal, S., Khan, I.U., Lee, J.W., 2022. A Transfer Learning-Based Approach to Detect Cerebral Microbleeds. Computers, Materials \& Continua 71.

Al-Masni, M.A., Kim, W.R., Kim, E.Y., Noh, Y., Kim, D.H., 2020. Automated detection of cerebral microbleeds in MR images: A two-stage deep learning approach. Neuroimage Clin 28, 102464.

Al-Masni, M.A., Kim, W.R., Kim, E.Y., Noh, Y., Kim, D.H., 2021. 3D Multi-Scale Residual Network Toward Lacunar Infarcts Identification From MR Images With Minimal User Intervention. IEEE Access 9, 11787-11797.

Balakrishnan, R., Valdes Hernandez, M.D.C., Farrall, A.J., 2021. Automatic segmentation of white matter hyperintensities from brain magnetic resonance images in the era of deep learning and big data - A systematic review. Comput Med Imaging Graph 88, 101867.

Ballerini, L., Booth, T., Valdés Hernández, M.d.C., Wiseman, S., Lovreglio, R., Muñoz Maniega, S., Morris, Z., Pattie, A., Corley, J., Gow, A., Bastin, M.E., Deary, I.J., Wardlaw, J., 2020. Computational quantification of brain perivascular space morphologies: Associations with vascular risk factors and white matter hyperintensities. A study in the Lothian Birth Cohort 1936. NeuroImage: Clinical 25, 102120.

Ballerini, L., Lovreglio, R., Hernández, M.d.C.V., Gonzalez-Castro, V., Maniega, S.M., Pellegrini, E., Bastin, M.E., Deary, I.J., Wardlaw, J.M., 2016. Application of the Ordered Logit Model to Optimising Frangi Filter Parameters for Segmentation of Perivascular Spaces. Procedia Computer Science 90, 61-67.

Ballerini, L., Lovreglio, R., Valdés Hernández, M.D.C., Ramirez, J., MacIntosh, B.J., Black, S.E., Wardlaw, J.M., 2018. Perivascular Spaces Segmentation in Brain MRI Using Optimal 3D Filtering. Sci Rep 8, 2132.

Barnes, S.R., Haacke, E.M., Ayaz, M., Boikov, A.S., Kirsch, W., Kido, D., 2011. Semiautomated detection of cerebral microbleeds in magnetic resonance images. Magn Reson Imaging 29, 844-852.

Bian, W., Hess, C.P., Chang, S.M., Nelson, S.J., Lupo, J.M., 2013. Computer-aided detection of radiation-induced cerebral microbleeds on susceptibility-weighted MR images. NeuroImage: Clinical 2, 282-290.

Billot, B., Bocchetta, M., Todd, E., Dalca, A.V., Rohrer, J.D., Iglesias, J.E., 2020. Automated segmentation of the hypothalamus and associated subunits in brain MRI. Neuroimage 223, 117287.

Boespflug, E.L., Schwartz, D.L., Lahna, D., Pollock, J., Iliff, J.J., Kaye, J.A., Rooney, W., Silbert, L.C., 2018. MR Imaging-based Multimodal Autoidentification of Perivascular Spaces (mMAPS): Automated Morphologic Segmentation of Enlarged Perivascular Spaces at Clinical Field Strength. Radiology 286, 632-642.

Boutinaud, P., Tsuchida, A., Laurent, A., Adonias, F., Hanifehlou, Z., Nozais, V., Verrecchia, V., Lampe, L., Zhang, J., Zhu, Y.C., Tzourio, C., Mazoyer, B., Joliot, M., 2021. 3D Segmentation of Perivascular Spaces on T1-Weighted 3 Tesla MR Images With a Convolutional Autoencoder and a U-Shaped Neural Network. Front Neuroinform 15, 641600.

Brown, R., Low, A., Markus, H.S., 2021. Rate of, and risk factors for, white matter hyperintensity growth: a systematic review and meta-analysis with implications for clinical trial design. Journal of Neurology, Neurosurgery & Psychiatry 92, 1271.

Caligiuri, M.E., Perrotta, P., Augimeri, A., Rocca, F., Quattrone, A., Cherubini, A., 2015. Automatic Detection of White Matter Hyperintensities in Healthy Aging and Pathology Using Magnetic Resonance Imaging: A Review. Neuroinformatics 13, 261-276.

Chen, H., Yu, L., Dou, Q., Shi, L., Mok, V.C.T., Heng, P.A., 2015. Automatic detection of cerebral microbleeds via deep learning based 3D feature representation. 2015 IEEE 12th International Symposium on Biomedical Imaging (ISBI), pp. 764-767.

Chen, Y., Villanueva-Meyer, J.E., Morrison, M.A., Lupo, J.M., 2019. Toward Automatic Detection of Radiation-Induced Cerebral Microbleeds Using a 3D Deep Residual Network. Journal of Digital Imaging 32, 766-772.

Cheng, A.L., Batool, S., McCreary, C.R., Lauzon, M.L., Frayne, R., Goyal, M., Smith, E.E., 2013. Susceptibility-weighted imaging is more reliable than T2*-weighted gradient-recalled echo MRI for detecting microbleeds. Stroke 44, 2782-2786.
Chesebro, A.G., Amarante, E., Lao, P.J., Meier, I.B., Mayeux, R., Brickman, A.M., 2021. Automated detection of cerebral microbleeds on T2*-weighted MRI. Scientific Reports 11, 4004.
Das, A.S., Regenhardt, R.W., Vernooij, M.W., Blacker, D., Charidimou, A., Viswanathan, A., 2019. Asymptomatic Cerebral Small Vessel Disease: Insights from Population-Based Studies. J Stroke 21, 121-138.
Debette, S., Markus, H.S., 2010. The clinical importance of white matter hyperintensities on brain magnetic resonance imaging: systematic review and meta-analysis. BMJ 341, c3666.
Descombes, X., Kruggel, F., Wollny, G., Gertz, H.J., 2004. An object-based approach for detecting small brain lesions: application to Virchow-Robin spaces. IEEE Transactions on Medical Imaging 23, 246-255.
Dou, Q., Chen, H., Yu, L., Shi, L., Wang, D., Mok, V.C., Heng, P.A., 2015. Automatic cerebral microbleeds detection from MR images via Independent Subspace Analysis based hierarchical features. 2015 37th Annual International Conference of the IEEE Engineering in Medicine and Biology Society (EMBC), pp. 7933-7936.
Dou, Q., Chen, H., Yu, L., Zhao, L., Qin, J., Wang, D., Mok, V.C., Shi, L., Heng, P., 2016. Automatic Detection of Cerebral Microbleeds From MR Images via 3D Convolutional Neural Networks. IEEE Transactions on Medical Imaging 35, 1182-1195.
Dubost, F., Adams, H., Bortsova, G., Ikram, M.A., Niessen, W., Vernooij, M., de Bruijne, M., 2019a. 3D regression neural network for the quantification of enlarged perivascular spaces in brain MRI. Medical Image Analysis 51, 89-100.
Dubost, F., Adams, H., Yilmaz, P., Bortsova, G., Tulder, G.v., Ikram, M.A., Niessen, W., Vernooij, M.W., Bruijne, M.d., 2020. Weakly supervised object detection with 2D and 3D regression neural networks. Medical Image Analysis 65, 101767.
Dubost, F., Bortsova, G., Adams, H.H.H., Ikram, M.A., Niessen, W.J., Vernooij, M.W., de Bruijne, M., 2017. GP-Unet: Lesion Detection from Weak Labels with a 3D Regression Network. MICCAI.
Dubost, F., Yilmaz, P., Adams, H., Bortsova, G., Ikram, M.A., Niessen, W., Vernooij, M., de Bruijne, M., 2019b. Enlarged perivascular spaces in brain MRI: Automated quantification in four regions. Neuroimage 185, 534-544.
Fazlollahi, A., Meriaudeau, F., Giancardo, L., Villemagne, V.L., Rowe, C.C., Yates, P., Salvado, O., Bourgeat, P., 2015. Computer-aided detection of cerebral microbleeds in susceptibility-weighted imaging. Comput Med Imaging Graph 46 Pt 3, 269-276.
Fazlollahi, A., Meriaudeau, F., Villemagne, V.L., Rowe, C.C., Yates, P., Salvado, O., Bourgeat, P., 2014. Efficient machine learning framework for computer-aided detection of cerebral microbleeds using the Radon transform. 2014 IEEE 11th International Symposium on Biomedical Imaging (ISBI), pp. 113-116.
Frangi, A.F., Niessen, W.J., Vincken, K.L., Viergever, M.A., 1998. Multiscale vessel enhancement filtering. In: Wells, W.M., Colchester, A., Delp, S. (Eds.), Medical Image Computing and Computer-Assisted Intervention — MICCAI'98. Springer Berlin Heidelberg, Berlin, Heidelberg, pp. 130-137.
Ghafaryasl, B., Lijn, F.v.d., Poels, M., Vrooman, H., Ikram, M.A., Niessen, W.J., Lugt, A.v.d., Vernooij, M., Bruijne, M.d., 2012. A computer aided detection system for cerebral microbleeds in brain MRI. 2012 9th IEEE International Symposium on Biomedical Imaging (ISBI), pp. 138-141.
Ghafoorian, M., Karssemeijer, N., Heskes, T., Bergkamp, M., Wissink, J., Obels, J., Keizer, K., de Leeuw, F.E., Ginneken, B.V., Marchiori, E., Platel, B., 2017. Deep multi-scale location-aware 3D convolutional neural networks for automated detection of lacunes of presumed vascular origin. Neuroimage Clin 14, 391-399.

Ghaznawi, R., Geerlings, M.I., Jaarsma-Coes, M.G., Zwartbol, M.H., Kuijf, H.J., van der Graaf, Y., Witkamp, T.D., Hendrikse, J., de Bresser, J., 2019. The association between lacunes and white matter hyperintensity features on MRI: The SMART-MR study. J Cereb Blood Flow Metab 39, 2486-2496.
González-Castro, V., Valdés Hernández, M.d.C., Armitage, P.A., Wardlaw, J.M., 2016. Automatic Rating of Perivascular Spaces in Brain MRI Using Bag of Visual Words. In: Campilho, A., Karray, F. (Eds.), Image Analysis and Recognition. Springer International Publishing, Cham, pp. 642-649.
González-Castro, V., Valdés Hernández, M.D.C., Chappell, F.M., Armitage, P.A., Makin, S., Wardlaw, J.M., 2017. Reliability of an automatic classifier for brain enlarged perivascular spaces burden and comparison with human performance. Clin Sci (Lond) 131, 1465-1481.
Greenberg, S.M., Vernooij, M.W., Cordonnier, C., Viswanathan, A., Al-Shahi Salman, R., Warach, S., Launer, L.J., Van Buchem, M.A., Breteler, M.M., 2009. Cerebral microbleeds: a guide to detection and interpretation. Lancet Neurol 8, 165-174.
Gregoire, S.M., Chaudhary, U.J., Brown, M.M., Yousry, T.A., Kallis, C., Jager, H.R., Werring, D.J., 2009. The Microbleed Anatomical Rating Scale (MARS): reliability of a tool to map brain microbleeds. Neurology 73, 1759-1766.
Groeschel, S., Chong, W.K., Surtees, R., Hanefeld, F., 2006. Virchow-Robin spaces on magnetic resonance images: normative data, their dilatation, and a review of the literature. Neuroradiology 48, 745-754.
Guan, H., Liu, M., 2022. Domain Adaptation for Medical Image Analysis: A Survey. IEEE Transactions on Biomedical Engineering 69, 1173-1185.
Guan, H., Wang, C., Tao, D., 2021. MRI-based Alzheimer's disease prediction via distilling the knowledge in multi-modal data. Neuroimage, 118586.
Guerrero, R., Qin, C., Oktay, O., Bowles, C., Chen, L., Joules, R., Wolz, R., Valdés-Hernández, M.C., Dickie, D.A., Wardlaw, J., Rueckert, D., 2018. White matter hyperintensity and stroke lesion segmentation and differentiation using convolutional neural networks. Neuroimage Clin 17, 918-934.
Hernandez Mdel, C., Piper, R.J., Wang, X., Deary, I.J., Wardlaw, J.M., 2013. Towards the automatic computational assessment of enlarged perivascular spaces on brain magnetic resonance images: a systematic review. J Magn Reson Imaging 38, 774-785.
Hong, J., Cheng, H., Zhang, Y.-D., Liu, J., 2019. Detecting cerebral microbleeds with transfer learning. Machine Vision and Applications 30, 1123-1133.
Hong, J., Wang, S.-H., Cheng, H., Liu, J., 2020. Classification of cerebral microbleeds based on fully-optimized convolutional neural network. Multimedia Tools and Applications 79, 15151-15169.
Hou, Y., Park, S.H., Wang, Q., Zhang, J., Zong, X., Lin, W., Shen, D., 2017. Enhancement of Perivascular Spaces in 7 T MR Image using Haar Transform of Non-local Cubes and Block-matching Filtering. Scientific Reports 7, 8569-8569.
Ibrahim, A.A., Ibrahim, Y.A., Darwish, E.A., Khater, N.H., 2019. Prevalence of cerebral microbleeds and other cardiovascular risk factors in elderly patients with acute ischemic stroke. Egyptian Journal of Radiology and Nuclear Medicine 50, 38.
Jung, E., Chikontwe, P., Zong, X., Lin, W., Shen, D., Park, S.H., 2019. Enhancement of Perivascular Spaces Using Densely Connected Deep Convolutional Neural Network. IEEE Access 7, 18382-18391.
Karimi, D., Dou, H., Warfield, S.K., Gholipour, A., 2020. Deep learning with noisy labels: Exploring techniques and remedies in medical image analysis. Medical Image Analysis 65, 101759.
Kuijf, H.J., Bresser, J.d., Biessels, G.J., Viergever, M.A., Vincken, K.L., 2011. Detecting cerebral microbleeds in 7.0 T MR images using the radial symmetry transform. 2011 IEEE International Symposium on Biomedical Imaging: From Nano to Macro, pp. 758-761.
Kuijf, H.J., Brundel, M., de Bresser, J., van Veluw, S.J., Heringa, S.M., Viergever, M.A., Biessels, G.J., Vincken, K.L., 2013. Semi-Automated Detection of Cerebral Microbleeds on 3.0 T MR Images. PLoS One 8, e66610.

Kuijf, H.J., de Bresser, J., Geerlings, M.I., Conijn, M.M.A., Viergever, M.A., Biessels, G.J., Vincken, K.L., 2012. Efficient detection of cerebral microbleeds on 7.0T MR images using the radial symmetry transform. Neuroimage 59, 2266-2273.
Li, T., Zou, Y., Bai, P., Li, S., Wang, H., Chen, X., Meng, Z., Kang, Z., Zhou, G., 2021. Detecting cerebral microbleeds via deep learning with features enhancement by reusing ground truth. Comput Methods Programs Biomed 204, 106051.
Lian, C., Liu, M., Zhang, J., Zong, X., Lin, W., Shen, D., 2018a. Automatic Segmentation of 3D Perivascular Spaces in 7T MR Images Using Multi-Channel Fully Convolutional Network. Proc Int Soc Magn Reson Med Sci Meet Exhib Int Soc Magn Reson Med Sci Meet Exhib 2018.
Lian, C., Zhang, J., Liu, M., Zong, X., Hung, S.C., Lin, W., Shen, D., 2018b. Multi-channel multi-scale fully convolutional network for 3D perivascular spaces segmentation in 7T MR images. Med Image Anal 46, 106-117.
Liu, C., Habib, T., Salimeen, M., Pradhan, A., Singh, M., Wang, M., Wu, F., Zhang, Y., Gao, L., Yang, G., Li, X., Yang, J., 2020a. Quantification of visible Virchow-Robin spaces for detecting the functional status of the glymphatic system in children with newly diagnosed idiopathic generalized epilepsy. Seizure 78, 12-17.
Liu, H., Rashid, T., Habes, M., 2020b. Cerebral Microbleed Detection Via Fourier Descriptor with Dual Domain Distribution Modeling. 2020 IEEE 17th International Symposium on Biomedical Imaging Workshops (ISBI Workshops), pp. 1-4.
Liu, S., Utriainen, D., Chai, C., Chen, Y., Wang, L., Sethi, S.K., Xia, S., Haacke, E.M., 2019. Cerebral microbleed detection using Susceptibility Weighted Imaging and deep learning. Neuroimage 198, 271-282.
Lu, D., Liu, J., MacKinnon, A.D., Tozer, D.J., Markus, H.S., 2021a. Prevalence and Risk Factors of Cerebral Microbleeds. Neurology 97, e1493.
Lu, S., Liu, S., Wang, S.H., Zhang, Y.D., 2021b. Cerebral Microbleed Detection via Convolutional Neural Network and Extreme Learning Machine. Front Comput Neurosci 15, 738885.
Lu, S., Lu, Z., Hou, X., Cheng, H., Wang, S., 2017. Detection of cerebral microbleeding based on deep convolutional neural network. 2017 14th International Computer Conference on Wavelet Active Media Technology and Information Processing (ICCWAMTIP), pp. 93-96.
Madan, C.R., 2021. Scan Once, Analyse Many: Using Large Open-Access Neuroimaging Datasets to Understand the Brain. Neuroinformatics.
Marcus, D.S., Fotenos, A.F., Csernansky, J.G., Morris, J.C., Buckner, R.L., 2010. Open access series of imaging studies: longitudinal MRI data in nondemented and demented older adults. Journal of Cognitive Neuroscience 22, 2677-2684.
Miller, K.L., Alfaro-Almagro, F., Bangerter, N.K., Thomas, D.L., Yacoub, E., Xu, J., Bartsch, A.J., Jbabdi, S., Sotiropoulos, S.N., Andersson, J.L.R., Griffanti, L., Douaud, G., Okell, T.W., Weale, P., Dragonu, I., Garratt, S., Hudson, S., Collins, R., Jenkinson, M., Matthews, P.M., Smith, S.M., 2016. Multimodal population brain imaging in the UK Biobank prospective epidemiological study. Nature Neuroscience 19, 1523-1536.
Mittal, S., Wu, Z., Neelavalli, J., Haacke, E.M., 2009. Susceptibility-weighted imaging: technical aspects and clinical applications, part 2. AJNR Am J Neuroradiol 30, 232-252.
Morrison, M.A., Payabvash, S., Chen, Y., Avadiappan, S., Shah, M., Zou, X., Hess, C.P., Lupo, J.M., 2018. A user-guided tool for semi-automated cerebral microbleed detection and volume segmentation: Evaluating vascular injury and data labelling for machine learning. NeuroImage: Clinical 20, 498-505.
Myung, M.J., Lee, K.M., Kim, H.G., Oh, J., Lee, J.Y., Shin, I., Kim, E.J., Lee, J.S., 2021. Novel Approaches to Detection of Cerebral Microbleeds: Single Deep Learning Model to Achieve a Balanced Performance. J Stroke Cerebrovasc Dis 30, 105886.
Nicola, Bluemke, E., Sundaresan, V., Jenkinson, M., Smith, S., Ana, 2021. Challenges for machine learning in clinical translation of big data imaging studies. arXiv pre-print server.

Park, S.H., Zong, X., Gao, Y., Lin, W., Shen, D., 2016. Segmentation of perivascular spaces in 7T MR image using auto-context model with orientation-normalized features. Neuroimage 134, 223-235.
Poels, M.M., Vernooij, M.W., Ikram, M.A., Hofman, A., Krestin, G.P., van der Lugt, A., Breteler, M.M., 2010. Prevalence and risk factors of cerebral microbleeds: an update of the Rotterdam scan study. Stroke 41, S103-106.
Rashid, T., Abdulkadir, A., Nasrallah, I.M., Ware, J.B., Liu, H., Spincemaille, P., Romero, J.R., Bryan, R.N., Heckbert, S.R., Habes, M., 2021. DEEPMIR: a deep neural network for differential detection of cerebral microbleeds and iron deposits in MRI. Scientific Reports 11, 14124.
Roerdink, J.B.T.M., Meijster, A., 2000. The Watershed Transform: Definitions, Algorithms and Parallelization Strategies. Fundamenta Informaticae 41, 187-228.
Roy, S., Jog, A., Magrath, E., Butman, J., Pham, D., 2015. Cerebral microbleed segmentation from susceptibility weighted images. SPIE.
Ruetten, P.P.R., Gillard, J.H., Graves, M.J., 2019. Introduction to Quantitative Susceptibility Mapping and Susceptibility Weighted Imaging. Br J Radiol 92, 20181016.
Sarmento, R.M., Vasconcelos, F.F.X., Filho, P.P.R., Wu, W., de Albuquerque, V.H.C., 2020. Automatic Neuroimage Processing and Analysis in Stroke-A Systematic Review. IEEE Rev Biomed Eng 13, 130-155.
Schwartz, D.L., Boespflug, E.L., Lahna, D.L., Pollock, J., Roese, N.E., Silbert, L.C., 2019. Autoidentification of perivascular spaces in white matter using clinical field strength T(1) and FLAIR MR imaging. Neuroimage 202, 116126.
Seghier, M.L., Kolanko, M.A., Leff, A.P., Jager, H.R., Gregoire, S.M., Werring, D.J., 2011. Microbleed detection using automated segmentation (MIDAS): a new method applicable to standard clinical MR images. PLoS One 6, e17547.
Sepehrband, F., Barisano, G., Sheikh-Bahaei, N., Cabeen, R.P., Choupan, J., Law, M., Toga, A.W., 2019. Image processing approaches to enhance perivascular space visibility and quantification using MRI. Scientific Reports 9, 12351.
Shams, S., Martola, J., Cavallin, L., Granberg, T., Shams, M., Aspelin, P., Wahlund, L.O., Kristoffersen-Wiberg, M., 2015. SWI or T2*: which MRI sequence to use in the detection of cerebral microbleeds? The Karolinska Imaging Dementia Study. AJNR Am J Neuroradiol 36, 1089-1095.
Sudre, C.H., Anson, B.G., Ingala, S., Lane, C.D., Jimenez, D., Haider, L., Varsavsky, T., Smith, L., Jäger, H.R., Cardoso, M.J., 2019. 3D multirater RCNN for multimodal multiclass detection and characterisation of extremely small objects. MIDL, pp. 447-456.
Tajudin, A.S., Sulaiman, S.N., Isa, I.S., Soh, Z.H.C., Karim, N.K.A., Shuaib, I.L., 2017a. Microbleeds detection using watershed-driven active contour. 2017 7th IEEE International Conference on Control System, Computing and Engineering (ICCSCE), pp. 320-324.
Tajudin, A.S., Sulaiman, S.N., Isa, I.S., Soh, Z.H.C., Tahir, N.M., Karim, N.K.A., Shuaib, I.L., 2017b. An improved watershed segmentation technique for microbleeds detection in MRI images. 2017 International Conference on Electrical, Electronics and System Engineering (ICEESE), pp. 11-16.
Uchiyama, Y., Abe, A., Muramatsu, C., Hara, T., Shiraishi, J., Fujita, H., 2015. Eigenspace template matching for detection of lacunar infarcts on MR images. J Digit Imaging 28, 116-122.
Uchiyama, Y., Asano, T., Hara, T., Fujita, H., Hoshi, H., Iwama, T., Kinosada, Y., 2009. CAD Scheme for differential diagnosis of lacunar infarcts and normal Virchow-Robin spaces on brain MR images. IFMBE 25, 126-128.
Uchiyama, Y., Asano, T., Kato, H., Hara, T., Kanematsu, M., Hoshi, H., Iwama, T., Fujita, H., 2012. Computer-aided diagnosis for detection of lacunar infarcts on MR images: ROC analysis of radiologists' performance. J Digit Imaging 25, 497-503.
Uchiyama, Y., Kunieda, T., Asano, T., Kato, H., Hara, T., Kanematsu, M., Iwama, T., Hoshi, H., Kinosada, Y., Fujita, H., 2008. Computer-aided diagnosis scheme for classification of lacunar infarcts and enlarged

Virchow-Robin spaces in brain MR images. 2008 30th Annual International Conference of the IEEE Engineering in Medicine and Biology Society, pp. 3908-3911.
Uchiyama, Y., Yokoyama, R., Ando, H., Asano, T., Kato, H., Yamakawa, H., Yamakawa, H., Hara, T., Iwama, T., Hoshi, H., Fujita, H., 2007a. Computer-aided diagnosis scheme for detection of lacunar infarcts on MR images. Acad Radiol 14, 1554-1561.
Uchiyama, Y., Yokoyama, R., Ando, H., Asano, T., Kato, H., Yamakawa, H., Yamakawa, H., Hara, T., Iwama, T., Hoshi, H., Fujita, H., 2007b. Improvement of Automated Detection Method of Lacunar Infarcts in Brain MR Images. 2007 29th Annual International Conference of the IEEE Engineering in Medicine and Biology Society, pp. 1599-1602.
van den Heuvel, T.L., Ghafoorian, M., van der Eerden, A., Goraj, B., Andriessen, T.M., ter Haar Romeny, B., Platel, B., 2015. Computer aided detection of brain micro-bleeds in traumatic brain injury. SPIE.
van den Heuvel, T.L.A., van der Eerden, A.W., Manniesing, R., Ghafoorian, M., Tan, T., Andriessen, T.M.J.C., Vande Vyvere, T., van den Hauwe, L., ter Haar Romeny, B.M., Goraj, B.M., Platel, B., 2016. Automated detection of cerebral microbleeds in patients with traumatic brain injury. NeuroImage: Clinical 12, 241-251.
Verdelho, A., Biessels, G.J., Chabriat, H., Charidimou, A., Duering, M., Godefroy, O., Pantoni, L., Pavlovic, A., Wardlaw, J., 2021. Cerebrovascular disease in patients with cognitive impairment: A white paper from the ESO dementia committee - A practical point of view with suggestions for the management of cerebrovascular diseases in memory clinics. Eur Stroke J 6, 111-119.
Wang, S., Jiang, Y., Hou, X., Cheng, H., Du, S., 2017. Cerebral Micro-Bleed Detection Based on the Convolution Neural Network With Rank Based Average Pooling. IEEE Access 5, 16576-16583.
Wang, S., Tang, C., Sun, J., Zhang, Y., 2019. Cerebral Micro-Bleeding Detection Based on Densely Connected Neural Network. Frontiers in Neuroscience 13.
Wang, X., Valdés Hernández Mdel, C., Doubal, F., Chappell, F.M., Piper, R.J., Deary, I.J., Wardlaw, J.M., 2016. Development and initial evaluation of a semi-automatic approach to assess perivascular spaces on conventional magnetic resonance images. J Neurosci Methods 257, 34-44.
Wang, Y., Catindig, J.A., Hilal, S., Soon, H.W., Ting, E., Wong, T.Y., Venketasubramanian, N., Chen, C., Qiu, A., 2012. Multi-stage segmentation of white matter hyperintensity, cortical and lacunar infarcts. Neuroimage 60, 2379-2388.
Wardlaw, J.M., Smith, E.E., Biessels, G.J., Cordonnier, C., Fazekas, F., Frayne, R., Lindley, R.I., O'Brien, J.T., Barkhof, F., Benavente, O.R., Black, S.E., Brayne, C., Breteler, M., Chabriat, H., DeCarli, C., de Leeuw, F.E., Doubal, F., Duering, M., Fox, N.C., Greenberg, S., Hachinski, V., Kilimann, I., Mok, V., van Oostenbrugge, R., Pantoni, L., Speck, O., Stephan, B.C.M., Teipel, S., Viswanathan, A., Werring, D., Chen, C., Smith, C., van Buchem, M., Norrving, B., Gorelick, P.B., Dichgans, M., Changes, S.R.V., 2013. Neuroimaging standards for research into small vessel disease and its contribution to ageing and neurodegeneration. Lancet Neurology 12, 822-838.
Yang, E., Gonuguntla, V., Moon, W.-J., Moon, Y., Kim, H.-J., Park, M., Kim, J.-H., 2021. Direct Rating Estimation of Enlarged Perivascular Spaces (EPVS) in Brain MRI Using Deep Neural Network. Applied Sciences 11, 9398.
Yokoyama, R., Zhang, X., Uchiyama, Y., Fujita, H., Hara, T., Zhou, X., Kanematsu, M., Asano, T., Kondo, H., Goshima, S., Hoshi, H., Iwama, T., 2007. Development of an Automated Method for the Detection of Chronic Lacunar Infarct Regions in Brain MR Images. IEICE Trans. Inf. Syst. 90-D, 943-954.
Zhang, J., Gao, Y., Park, S.H., Zong, X., Lin, W., Shen, D., 2016a. Segmentation of Perivascular Spaces Using Vascular Features and Structured Random Forest from 7T MR Image. Mach Learn Med Imaging 10019, 61-68.
Zhang, J., Gao, Y., Park, S.H., Zong, X., Lin, W., Shen, D., 2017. Structured Learning for 3-D Perivascular Space Segmentation Using Vascular Features. IEEE Trans Biomed Eng 64, 2803-2812.

Zhang, Y.-D., Hou, X.-X., Chen, Y., Chen, H., Yang, M., Yang, J., Wang, S.-H., 2018a. Voxelwise detection of cerebral microbleed in CADASIL patients by leaky rectified linear unit and early stopping. Multimedia Tools and Applications 77, 21825-21845.
Zhang, Y.-D., Zhang, Y., Hou, X.-X., Chen, H., Wang, S.-H., 2018b. Seven-layer deep neural network based on sparse autoencoder for voxelwise detection of cerebral microbleed. Multimedia Tools and Applications 77, 10521-10538.
Zhang, Y.D., Hou, X.X., Lv, Y.D., Chen, H., Zhang, Y., Wang, S.H., 2016b. Sparse Autoencoder Based Deep Neural Network for Voxelwise Detection of Cerebral Microbleed. 2016 IEEE 22nd International Conference on Parallel and Distributed Systems (ICPADS), pp. 1229-1232.